\documentclass[11pt,a4paper]{article}

\pdfoutput=1

\usepackage{jheppub}
\usepackage{amsmath}
\usepackage{amsfonts}
\usepackage{graphicx}
\usepackage{amssymb}
\usepackage{amssymb}
\usepackage{latexsym}
\usepackage{array}
\usepackage{bbold}
\usepackage[normalem]{ulem}
\usepackage{xcolor}
\usepackage{slashed}
\usepackage{cancel}

\newcommand{\bea}{\begin{eqnarray}}
\newcommand{\eea}{\end{eqnarray}}
\newcommand{\be}{\begin{equation}}
\newcommand{\ee}{\end{equation}}

\usepackage{xcolor}
\definecolor{mygreen}{HTML}{006E28}

\title{Oscillons and bubbles in $Q$-ball dynamics}

\author[a]{D. Canillas Mart\'inez}
\affiliation[a]{IUFFyM, University of Salamanca, Plaza de la Merced 1, 37008 - Salamanca, Spain}
\emailAdd{dnl.canillas@usal.es}

\author[b]{P. Dorey}
\affiliation[b]{Department of Mathematical Sciences, Durham University, Durham DH1 3LE, UK}
\emailAdd{p.e.dorey@durham.ac.uk}

\author[c]{T. Roma\'nczukiewicz}
\affiliation[c]{Faculty of Theoretical Physics, Astronomy and Applied Computer Science, Jagiellonian University, Krak\'ow, Poland}
\emailAdd{tomasz.romanczukiewicz@uj.edu.pl}

\author[d]{Paul M. Saffin}
\affiliation[d]{School of Physics and Astronomy, University of Nottingham, University Park, Nottingham NG7
2RD, United Kingdom}
\emailAdd{paul.saffin@nottingham.ac.uk}

\author[c]{K. S\l awi\'nska}
\emailAdd{katarzyna.slawinska@uj.edu.pl}

\author[c,g]{A. Wereszczy\'nski}
\affiliation[g]{International Institute for Sustainability with Knotted Chiral Meta Matter (WPI-SKCM$^{\;2}$), Hiroshima University, 1-3-1 Kagamiyama, Higashi-Hiroshima,Hiroshima 739-8526, Japan}
\emailAdd{andrzej.wereszczynski@uj.edu.pl}

\abstract{
We show that, in the thin-wall regime, $Q$-ball--anti-$Q$-ball collisions reveal chaotic behaviour. This is explained by the resonant energy transfer mechanism triggered by the internal modes hosted by the $Q$-balls and by   the existence of {\it ephemeral} states, that is unstable, sometimes even short-lived, field configurations that appear as intermediate states. The most important examples of such states are the {\it bubble} of the false broken vacuum, which as intermediate states govern the $QQ^*$ annihilation, and the {\it charged oscillons}. 

The usually short-lived bubble can be dynamically temporarily stabilized, which explains their importance in the dynamics of $Q$-balls. This happens due to the excitation of massless Goldstone modes, which, exerting pressure on the bubble boundaries or being trapped as bound modes, prevent the bubble from collapsing. 
}

\begin{document}

\maketitle

%%%%%%%%%%%%%%%%%%%%%%%%%%%%%%
\section{Introduction}
%%%%%%%%%%%%%%%%%%%%%%%%%%%%%%
$Q$-balls are nontopological solitons, that is, particle-like non-perturbative excitations, which carry a non-zero value of a Noether charge arising from a global $U(1)$ symmetry \cite{Friedberg:1976me, Coleman:1985ki}. This charge corresponds to a target-space rotation of the complex field and can stabilize the soliton against small perturbations.

$Q$-balls have found applications in various areas of contemporary physics, especially in the cosmological and astrophysical contexts, where they are considered as candidates for dark matter \cite{Kusenko:1997si, Dine:2003ax, Enqvist:2003gh, Cardoso:2019rvt}. Although not existing in the standard model, they do show up in its supersymmetric extensions. Specifically, they can be produced by a fragmentation of Affleck-Dine condensates, leading to a characteristic modification of the stochastic gravitational wave background \cite{Kusenko:2008zm}. If coupled to gravity, the $Q$-balls transmute into boson stars \cite{Liebling:2012fv, Cardoso:2019rvt}. With some extension, they can also reproduce galactic rotation curves \cite{Mourelle:2025ilv}. 

Understanding collisions of $Q$ balls can be of fundamental importance for their cosmological application. Once formed in the fragmentation of the homogeneous condensate of the complex field, $Q$-balls collide. This may alter their final density and therefore affect their application as a candidate for the dark matter content of the Universe. 

However, although the properties of single $Q$ balls are very well described (e.g., static properties, stability \cite{Friedberg:1976me, Coleman:1985ki, Lee:1991ax, Multamaki:1999an}, and the spectrum of small perturbations \cite{Ciurla:2024ksm}), their dynamics are far from being fully understood. Contrary to the domain wall case, interactions and, in particular, collisions of $Q$-balls have not yet been investigated in a comprehensive manner, see \cite{Battye:2000qj, Axenides:1999hs, Multamaki:2000qb, Multamaki:2000ey, Bowcock:2008dn} for pioneering works. Nevertheless, several very interesting effects have been reported; see, for example, the charge-swapping phenomenon \cite{Copeland:2014qra}, superradiance \cite{Saffin}, \cite{Cardoso:2023dtm}, or chaotic patterns in the final state formation \cite{Alonso-Izquierdo:2025iet}. 

\vspace*{0.2cm}

A generic source of the complexity of soliton dynamics is the {\it resonant energy transfer mechanism} originating in the fact that solitons host bound modes \cite{Campbell:1983xu, Sugiyama:1979mi}. This is also the case for $Q$-balls. However, the specific main reason for the plethora of scenarios in $Q$-ball dynamics is the simultaneous existence of other soliton-like excitations. The first example is an {\it oscillon}, that is, a quasi-periodic solitonic solution  \cite{Bogolyubsky:1976nx, Gleiser:1993pt, Copeland:1995fq}. Such an oscillon can be easily found in the real-field truncation of the complex field theory. Thus, it does not carry the $U(1)$ charge. It is not an ultimately stable state, but all the time emits (often extremely little) radiation \cite{Graham:2006xs, Salmi:2012ta, Fodor:2008du, Fodor:2009kf, Zhang:2020bec, Olle:2020qqy, vanDissel:2023zva}. However, it is not the only oscillon in the complex scalar theory. 

It is now well understood that $Q$-balls and oscillons are very closely related, with the properties of one object anchored in the other \cite{Blaschke:2025anm} (also \cite{Blaschke:2024dlt}). This relationship has a profound impact on dynamics. E.g., it has recently been shown that the charge-swapping state is an excited oscillon rather than a bound state of a single {\it individual} $Q$-ball and anti-$Q$-ball \cite{Alonso-Izquierdo:2025iet}. In fact, there is a whole family of long-lived states, called {\it polarized $Q$-balls}, which in a smooth way join the $Q$-ball solution with the oscillon solution, passing through excited $Q$-balls and {\it charged oscillons}. 

The second example is the {\it bubble} of the false vacuum. It is an unstable, fast-decaying configuration with zero $U(1)$ charge, which appears when the potential has an additional local minimum at $|\phi|\neq 0$. The bubble can be understood as a kink-antikink configuration immersed in the complex field theory. It rapidly decays as a result of the pressure that arises between the true vacuum (outside) and the false vacuum (inside).

The third soliton-like state is the {\it sphaleron} or {\it $Q$-sphaleron}. While the bubble is a region of false vacuum within true vacuum, the sphaleron is a saddle-point solution comprising true vacuum within the false vacuum. Once perturbed, it can decay, e..g, to an oscillon.  

These three types of non-perturbative excitations can be gathered under the common name of {\it ephemeral states}, that is, unstable, long- or even short-lived configurations that, contrary to the $Q$ balls, are absent in the final state of the evolution. Consequently, one could think that they are not too important for the evolution of $Q$-balls. We show that this is definitely not the case. The dynamics of $Q$-balls, and especially, the $QQ^*$ collisions, is very strongly affected by the existence of such ephemeral states. This is related to the fact that, during collisions, the ephemeral states are easily created and even those that are unstable can be temporarily stabilized by hosting additional excitations such as massless Goldstone modes in the case of the bubble. 

%%%%%%%%%%%%%%%%%%%%%%%%%%%%%%
\section{Complex $\phi^6$ model and $Q$-balls}
\begin{figure}%[h!]
 \centering
     \includegraphics[width=1.00\textwidth]{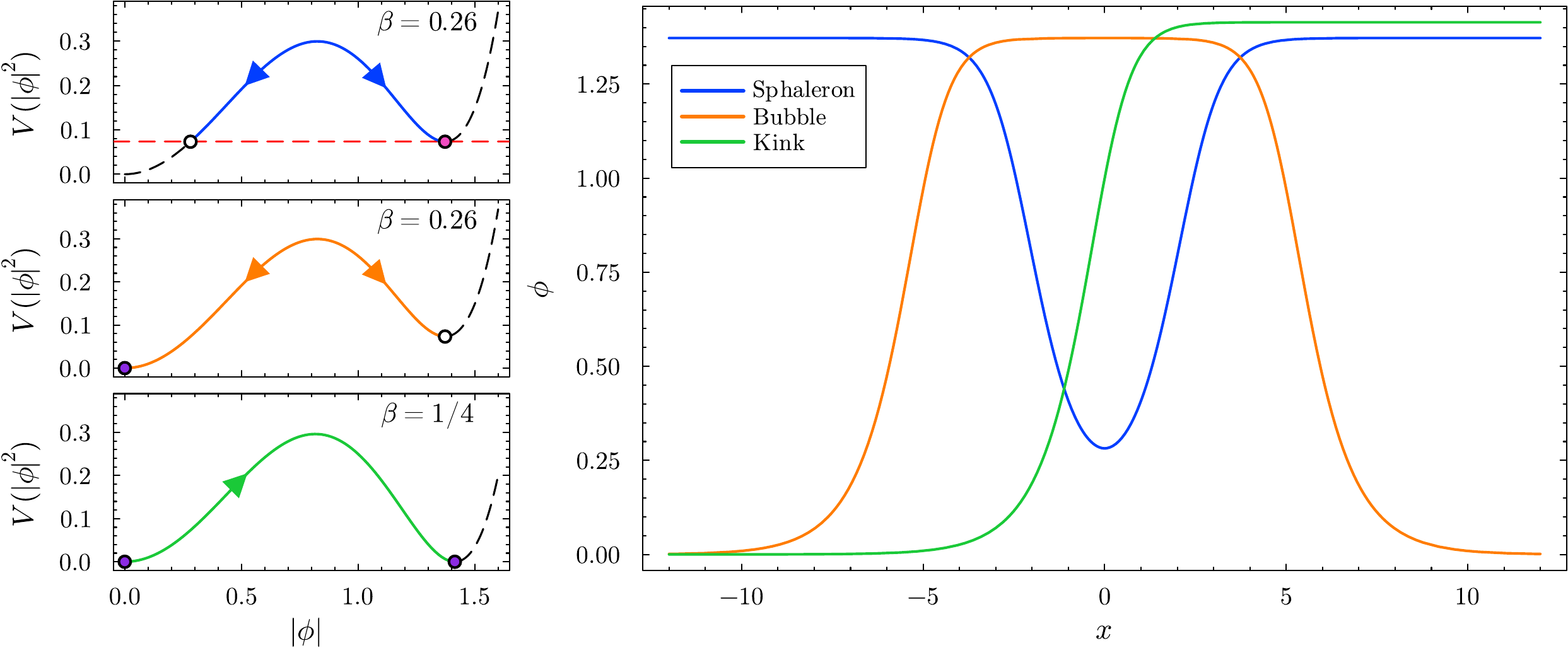}
     \caption{The potential $V$ (dashed curve) for $\beta = 0.26$ for which the false vacuum exists together with the charge-less states: the kink, sphaleron and bubble with a size parameter of $a=5$. Two arrows means that the solution goes there and back, e.g., for the bubble it starts and ends at the true vacuum, reaching the false vacuum in between.}
     \label{fig:V}   
 \end{figure}

%%%%%%%%%%%%%%%%%%%%%%%%%%%%%%
We begin with the simplest model supporting $Q$-balls and consider the complex $\phi^6$ theory in (1+1) dimensions defined by the following Lagrange density
\begin{equation}
    \mathcal{L} = \partial_\mu\phi\partial^\mu\phi^*-V(|\phi|),
\end{equation}
where the potential
\begin{equation}
  V(|\phi|)=|\phi|^2-|\phi|^4+\beta |\phi|^6\,
    \label{lag}
\end{equation}
contains, after suitable rescaling, only one nontrivial free parameter $\beta$. We assume that $\beta \geq 1/4$, which guarantees that $V(\phi) \geq 0$ with the vacuum at $\phi=0$. Importantly, there are two qualitatively distinct regimes. For $\beta > 1/3$ the potential is a monotonic function of the modulus. For $\beta \in \left(\frac{1}{4}, \frac{1}{3} \right)$ the potential has an additional local minimum, see Fig. \ref{fig:V}, a {\it false vacuum} at 
\begin{equation}
|\phi_\textrm{min}|^2=\frac{1+\sqrt{1-3\beta}}{3\beta},
\end{equation}
for which the value of the potential is  
\begin{equation}
    V\left(|\phi_\textrm{min}|\right) = 
    -\frac{\left(1+\sqrt{1-3\beta}\right)\left(1-6\beta+\sqrt{1-3\beta}\right)}{{27\beta^2}}.
    \label{eq:gap}
\end{equation}
At $\beta=\frac{1}{4}$ the false vacuum becomes a second, degenerate true vacuum that spontaneously breaks the U(1) symmetry, see Fig. \ref{fig:V}. 

\vspace*{0.2cm}

The field equation reads
\begin{equation}
\phi_{tt}-\phi_{xx}+(1-2|\phi|^2+3\beta |\phi|^4) \phi=0.
\label{field_eq}
\end{equation}
Due to the global $U(1)$ invariance, there is a conserved Noether charge 
\begin{equation}
Q=\int_{-\infty}^\infty j^0 \, dx, \;\;\;     j_\mu =i \left( \phi^* \partial_\mu \phi -  \phi \partial_\mu \phi^* \right),
\end{equation}
which is interpreted as a particle number. 

\begin{figure}%[h!]
 \centering
     \includegraphics[width=1.00\textwidth]{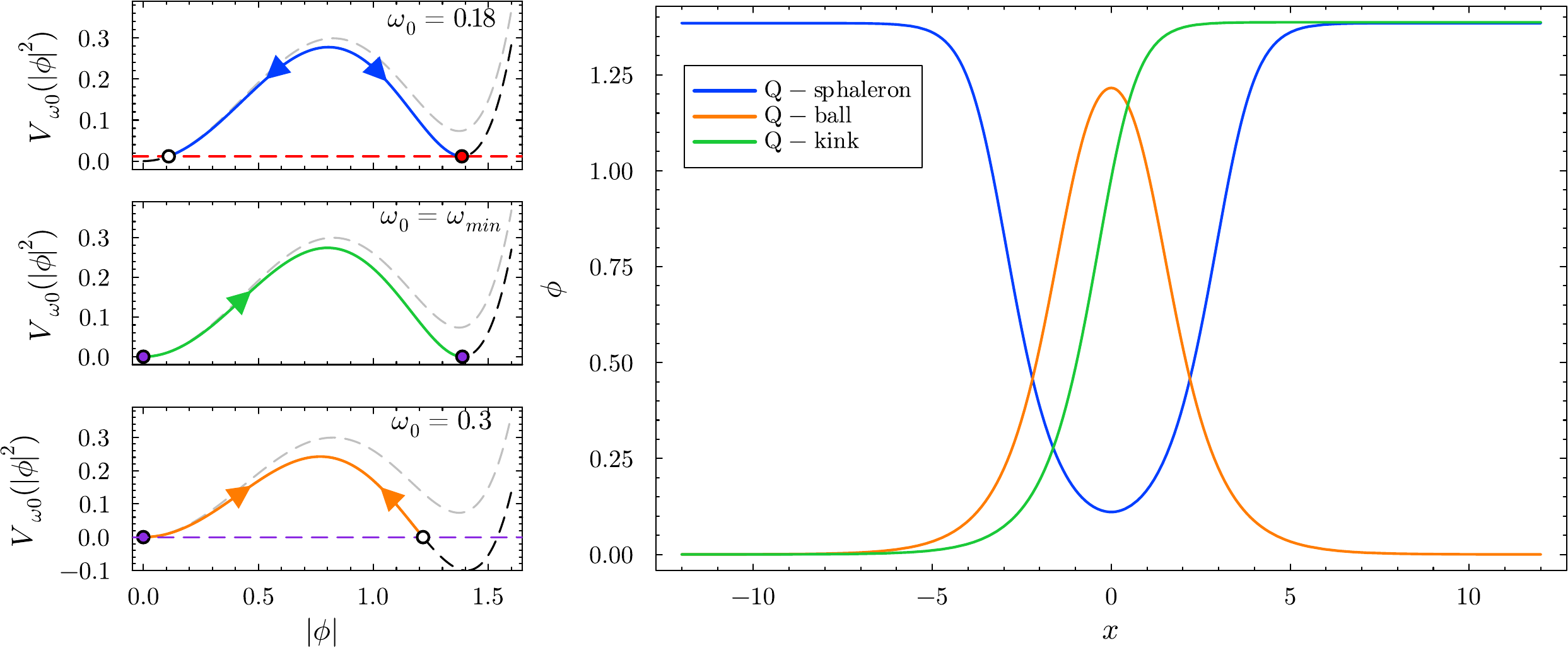}
     \caption{The effective potential $V_{\omega_0}$ for $\beta = 0.26$ and a few $\omega_0$ together with the charged states: the $Q$-ball, $Q$-kink and $Q$-sphaleron. }
     \label{fig:Veff}   
 \end{figure}

For $\beta \geq 1/4$, there exist localized, stationary solitonic-like solutions carrying a non-zero amount of the $U(1)$ charge, called {\it $Q$-balls}. They are given in the following form
\begin{equation}
   \phi_Q(x,t)= f_{\omega_0}(x) e^{i\omega_0 t}, 
\label{Q-solution}
\end{equation}
where $f_{\omega_0}$ is a profile associated with the internal rotation with frequency $\omega_0$. Such stationary states can be viewed as ``static" solutions joining the zeros of the effective potential $V_{\omega}(f^2)$
\begin{equation}
    V_{\omega_0} (f^2)\equiv - |\partial_t \phi|^2 +V(|\phi|) = (1-\omega^2_0) f_{\omega_0}^2-f_{\omega_0}^4+\beta f_{\omega_0}^6, 
    \end{equation}
    see Fig. \ref{fig:Veff}, orange curve. 
The finite-energy condition imposes that $f_{\omega_0}(x=\pm \infty)=0$. The other zero of the $V_{eff}$ occurs at the center of the $Q$-ball. In this picture, the $Q$-ball looks like the real $\phi^6$ sphaleron stabilized by the field-space rotation. 
 
Specifically, for our potential, the profile of the $Q$-ball reads
\begin{equation}
  f_{\omega_0}(x)=\frac{\sqrt{2}\epsilon }{\sqrt{1+\sqrt{1-4\beta\epsilon ^2}\cosh(2\epsilon x)}}\,,
\label{solution}
\end{equation}
where $\epsilon =\sqrt{1-\omega_0^2}$. These $Q$-balls exist for $\omega_0 \in (\omega_{min},\omega_{max})$, where 
\begin{equation}
    \omega_{min}=\sqrt{1-\frac{1}{4\beta}},
\end{equation}
while $\omega_{max}=1$, which equals the mass of small perturbations in the vacuum $\phi_{v}=0$. Importantly, not all $Q$-balls are stable - only those that obey the following condition  
\begin{equation}
    \frac{\omega_0}{Q} \frac{dQ}{d\omega_0} <0. 
\end{equation}
The others decay into stable counterparts with the same $U(1)$ charge. Here, the energy and the charge of the $Q$-balls are
\begin{equation}
    E (\omega_0) = \frac{4\omega_0 {\epsilon } + Q (4\beta-1 + 4 \beta \omega_0^2 )}{8\omega_0 \beta},\;\;\;\; 
Q(\omega_0) = \frac{4\omega_0}{\sqrt \beta}{\rm arctanh}\left(\frac{1-\sqrt{1-4\beta {\epsilon }^2}}{2{\epsilon } \sqrt \beta}\right)\,.
\label{EQ}
\end{equation}
 
There are two qualitatively distinct regimes. The first is the so-called {\it thick-wall} regime occurring for frequencies close to the mass threshold $\omega_{max}$. Such $Q$-balls have small energy and charge localized in their bulk. In this regime $Q$-balls are weakly bound multi-particle states of the mesons of the complex field. In fact, they are typically unstable due to quantum corrections. 

The second regime is the {\it thin-wall} regime. This happens when $\omega_0 \to \omega_{min}$. Here, the $Q$-balls are heavy and carry a large value of the charge. Importantly, the energy density is localized at the surface. Thus, in (1+1) dimensions they reveal a non-trivial internal structure with two peaks in the energy density concentrated in the vicinity of the surface. From the point of view of the effective potential, the thin-wall regime corresponds to the appearance of two degenerate minima at $f=0$ and $f=\frac{1}{\sqrt{2\beta}}$. The profile of the $Q$-ball looks like it is composed of a kink of $V_{\omega}$ (interpolating between $f=0$ and $f=\frac{1}{\sqrt{2\beta}}$) and an antikink (interpolating between $f=\frac{1}{\sqrt{2\beta}}$ and $f=0$), which are separated by an increasing plateau with $f=\frac{1}{\sqrt{2\beta}}$. 

$Q$-balls, as other solitons, host bound modes that can be found as solutions of the linear perturbation problem. The consistency of the linear perturbation of the oscillating solution requires the inclusion of two components $\eta_{1,2}$ with frequencies $\omega_0\pm \rho$. Thus, in the case of a $Q$-ball solution rotating with $\omega_0$
\begin{equation}
    \delta \phi (x,t) = \eta_1(x)e^{i(\omega_0+\rho)t} +\eta_2(x)e^{i(\omega_0-\rho)t}.
\end{equation}
If the frequencies of both components are below the mass threshold, we find a genuine bound mode. However, it is possible that one of these frequencies is above the mass threshold. This leads to a half-bound mode, that is, a quasi-normal mode also referred to as the Feshbach resonance. There is also a continuous family of scattering modes, where both $\omega_0\pm \rho$ are larger than the mass threshold.

\begin{figure}%[!h] 
    \centering
    \includegraphics[width=1.1\linewidth]{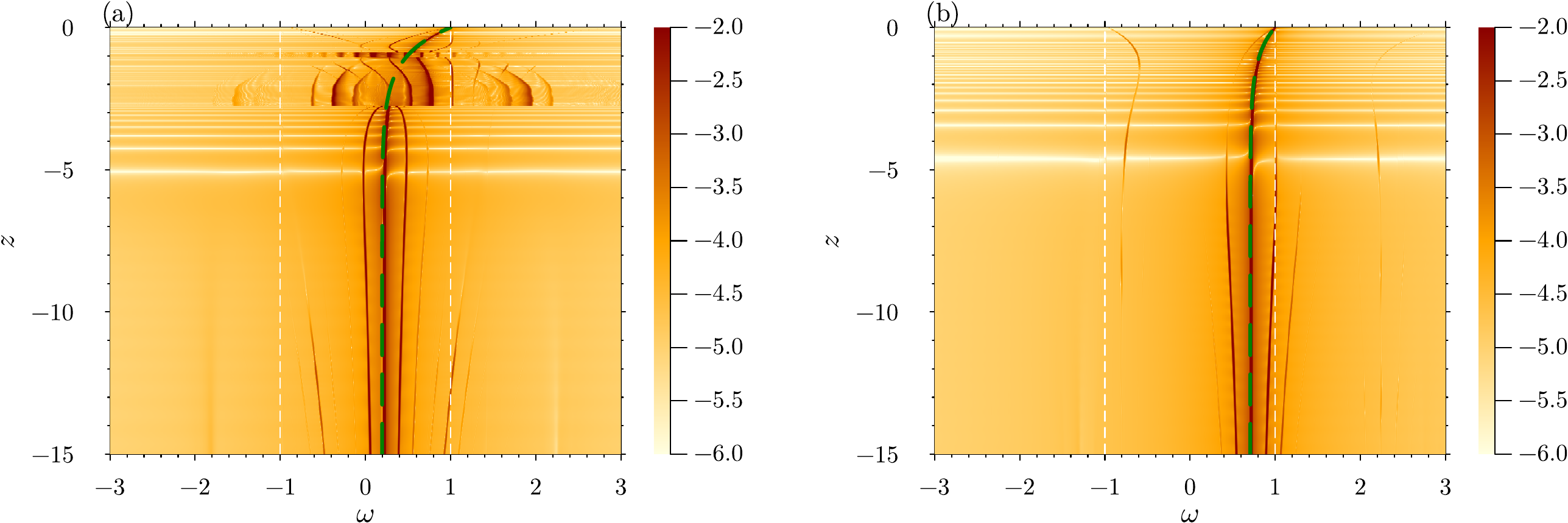}
    \caption{Modes in the power spectrum of a perturbed $Q$-ball as a function of $z=\ln \left( \frac{\omega_0-\omega_{min}}{1-\omega_{min}} \right)$, such that $z\to-\infty$ corresponds to thin-wall Q-balls and $z\to0$ to thick-wall Q-balls. The green dashed line is $\omega=\omega_0$, while the vertical white dashes lines are the mass threshold. Left: $\beta=0.26$, such that $V(|\phi|)$ has a false vacuum and $\omega_{min}\simeq0.196$. Right: $\beta=0.5$, such that $V(|\phi|)$ is monotonic and $\omega_{min}=1/\sqrt{2}$. $\omega$ is the frequency in the power spectrum.}
    \label{fig:QB-modes}
\end{figure}

Since the vibrational modes are known to play a distinguished role in solitonic collisions due to their participation in the resonant energy transfer mechanism, it is important to know how the spectral structure varies with $\omega_0$ for different $\beta$. This is plotted in Fig. \ref{fig:QB-modes}, where we present the modes visible in the power spectrum of perturbed (squeezed) $Q$-balls. To implement the squeezing of a Q-ball, following \cite{Bowcock:2008dn}, we start with a Q-ball solution and then construct a new configuration (with the same charge) 
 \begin{equation}
    \phi_\lambda(x,t)=\sqrt{\lambda}f_{\omega_0}(\lambda x)\,\qquad\textrm{with } \lambda=1.05\,.
\end{equation}
To find the power spectrum we then perform an FFT of the evolved field for $T=1000$. Then we applied a gaussian window with $4\sigma=T$ (in order to reduce the noise) to the field values at the centre and performed the standard Fast Fourier Transform (FFT).

\begin{figure}%[!h] 
    \centering
    \includegraphics[width=0.95\linewidth]{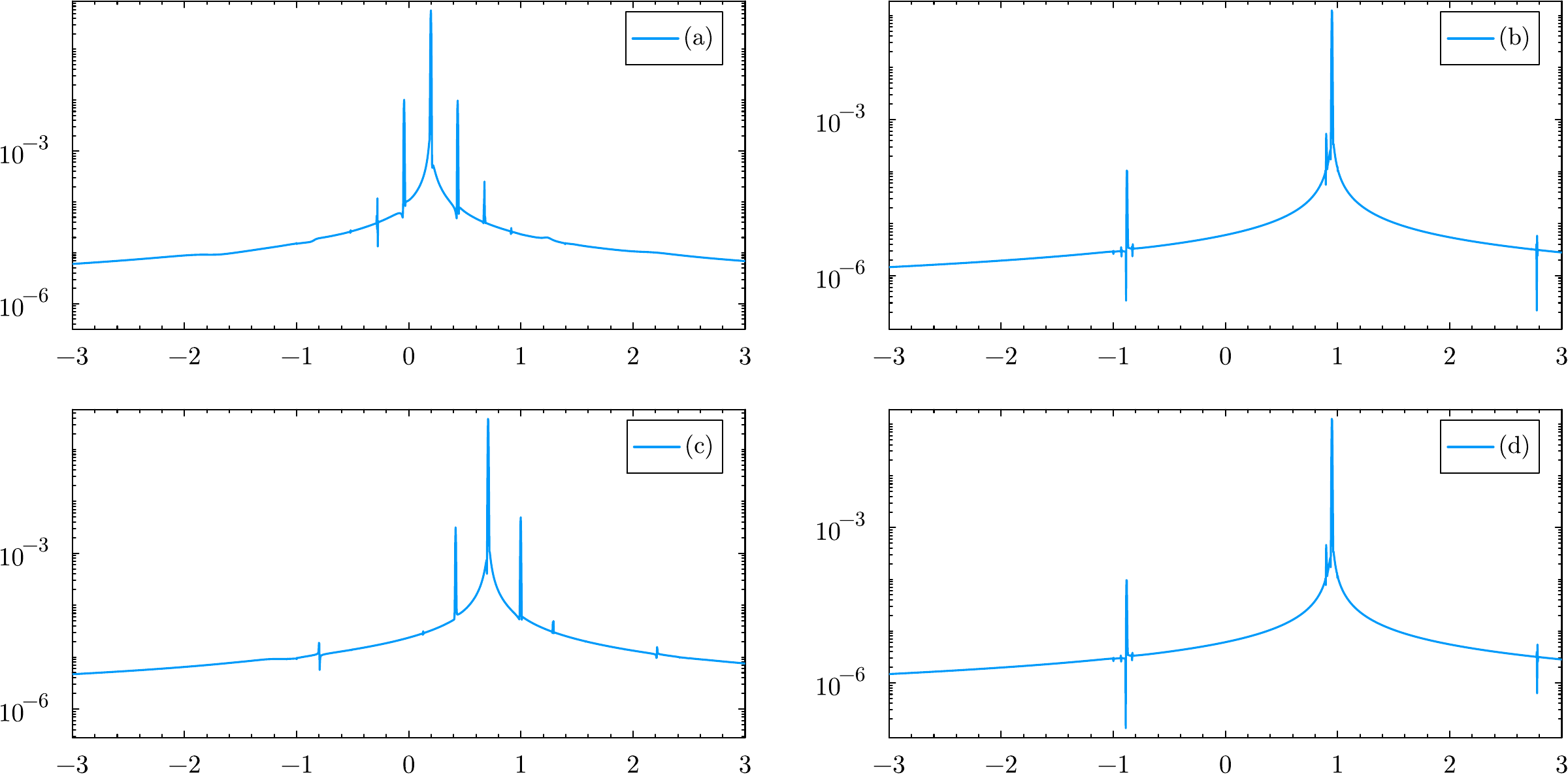}
    \caption{Modes in the power spectrum of a perturbed $Q$-ball: a) $\beta=0.26$ and $\omega_0=\omega_{min}+0.001 = 0.197$; b)  $\beta=0.26$ and $\omega_0=0.95$; c) $\beta=0.5$ and $\omega_0=\omega_{min}+0.001 = 0.708$; d) $\beta=0.5$ and $\omega_0=0.95$.}
    \label{fig:QB-modes-2}
\end{figure}

We found that in the thick-wall regime, the spectral structure is almost trivial. The $Q$-balls do not host any well-pronounced bound modes as the fundamental frequency $\omega_0$ is too close to the mass threshold. See also Fig. \ref{fig:QB-modes-2} where we plot the power spectrum of the squeezed $Q$-ball for $\beta=0.26$ and $\beta=0.5$ with $\omega_0=0.95$.  In the spectrum, we see only the peak at $\omega=\omega_0$ and a small peak from one component of a quasinormal mode. This suggests that in the thick-wall limit, there are no modes that could participate in the resonant energy transfer, which would trigger chaotic behaviour in the collisions. 

However, in the thin-wall regime, the $Q$-balls reveal a very well-pronounced bound mode, with both components below the mass threshold, see region with $z<-10$ in Fig. \ref{fig:QB-modes} and Fig.  \ref{fig:QB-modes-2}, panels a) and c). Here $z=\ln \left( \frac{\omega_0-\omega_{min}}{1-\omega_{min}} \right)$. For $\omega_0$ closer to $\omega_{min}$ and smaller $\beta$ there are more bound modes. Definitely, such $Q$-balls should possess more complex dynamics with a well-visible chaotic pattern in the final state formation. 

In Fig. \ref{fig:QB-modes}, left panel, we see a clearly visible region with intermediate $\omega_0$, where there is no main peak at $\omega_0$. This corresponds to the unstable region of $Q$-balls. 
%%%%%%%%%%%%%%%%%%%%%%%%%%%%%%
\section{The oscillons and polarized $Q$-balls}
%%%%%%%%%%%%%%%%%%%%%%%%%%%%%% 
The complex $\phi^6$ model can be consistently reduced to a version where $\phi \in \mathbb{R}$. Such a theory is known to support oscillons, which are localized, time-dependent quasi-periodic solutions. Although long-lived, the oscillons are not stable solutions. All the time, they emit radiation and eventually decay to a vacuum. The real-valued oscillons have zero $U(1)$ charge. Of course, they can be trivially rotated by a constant phase to a complex solution. 

Such a real-valued oscillon can easily be perturbed. If the perturbation probes the real component of the field, we typically get a modulated oscillon, which can also be understood as a pair of two fundamental, unperturbed oscillons forming a bound state \cite{Blaschke:2024dlt}. If the perturbation excites the imaginary component of the field, we find a {\it charged oscillon}, that is, a state where the oscillon hosts some non-zero density of the $U(1)$ charge. In fact, there is a whole family of long-lived solutions, called {\it polarized $Q$-balls}, which, passing through perturbed $Q$-balls and charged oscillons, interpolate in a smooth way between $Q$-balls and real-valued oscillons \cite{Blaschke:2025anm}. 

The existence of such charged oscillons is manifested in the charge-swapping solution \cite{Alonso-Izquierdo:2025iet}. This is another long-lived state in which the charge density quasi-periodically flows between different spatial regions \cite{Copeland:2014qra}. If the net charge vanishes, this solution eventually relaxes to a real oscillon \cite{Copeland:2014qra}. 

It is worth underlining that the charged oscillon, or in general the polarized $Q$-ball, is a very long-lived and very easy-to-excite state. This is because it is based on the $Q$-ball (or the oscillon) and its (half-)bound modes. It is therefore plausible to expect them to be created in collisions of the $Q$-balls. 

%%%%%%%%%%%%%%%%%%%%%%%%%%%%%%
\section{The $Q$-kinks and $Q$-sphalerons}
%%%%%%%%%%%%%%%%%%%%%%%%%%%%%%
The thin-wall limit is directly related to the existence of another stationary solution, i.e., the $Q$-kink ($Q$-antikink)
\begin{equation}
    \phi_{QK}(x,t)=\frac{1}{2\sqrt{\beta}} \sqrt{1\pm \tanh \left( \frac{x}{2\sqrt{\beta}}\right)} \, e^{i\omega_{min}t}.
\end{equation}
This solution interpolates between the two degenerate minima of the effective potential $V_{\omega_0}$, see Fig. \ref{fig:Veff}, green curve. It rotates in field space with the single frequency $\omega_{min}$ and carries an infinite amount of energy and $U(1)$ charge. However, when combined, the $Q$-kink and the $Q$-antikink form a finite-energy solution that tends to the thin wall $Q$ ball as $\omega_0 \to \omega_{min}$ (similar solutions also exist in non-flat target spaces \cite{AlonsoIzquierdo:2024chs}). In other words, the $Q$-ball in the thin-wall regime can be treated as a bound state of a $Q$-kink and $Q$-antikink
\begin{equation}
\phi_{Q}(x,t)\approx\phi_{QK}(x+x_0,t)+\phi_{QAK}(x-x_0,t)-\sqrt{\frac{2}{\beta}}e^{i\omega t}.
\end{equation}
To find this relation we compare the leading term of the profile for $x\to\infty$ 
\begin{equation}
    |\phi_Q(x,t)|\approx\frac{1}{\sqrt{2\beta}}\exp\left(-\frac{x-x_0}{2\sqrt{\beta}}\right)
\end{equation}
with the expansion of the exact Q-ball profile
\begin{equation}
    f_Q(x)\approx 2\epsilon(1-4\beta\epsilon^2)^{-1/4}e^{-\epsilon x}.
\end{equation}
Then, taking into account that around $\omega_0=\omega_{min}$
\begin{equation}
    \epsilon \approx\frac{1}{2\sqrt{\beta}} 
- 2 \omega_{min}\sqrt{\beta} (\omega_0 - \omega_{min})
\end{equation}
we find that
\begin{equation}
    \frac{1}{\sqrt{2\beta}}\exp\left(\frac{x_0}{2\sqrt{\beta}}\right)=2\epsilon(1-4\beta\epsilon^2)^{-1/4}
\end{equation}
which after some algebra gives 
\begin{equation}
    x_0 \approx\sqrt{\beta} \log\left( \frac{8 \beta \epsilon^2}{\sqrt{1 - 4 \epsilon^2 \beta}} \right) \approx \frac{\sqrt{\beta}}{4}\log\left((4\beta-1)\beta(\omega-\omega_{min})^2\right)\,.
\label{eq:a_for_phiQ}
\end{equation}
Note that we also neglected the $\mathcal{O}(\omega_0 - \omega_{min})$ correction in the exponent.

The constant value of the field in the plateau region corresponds to the value of the field at the second minimum of $V_{\omega_0}$. 

In the limit where $\beta=\frac{1}{4}$, $\omega_{min}=0$ and the $Q$-kink solutions become the static kinks of the (real) $\phi^6$ model, see Fig. \ref{fig:V}, green curve.
\begin{equation}
    \phi_{K}(x)= \sqrt{1\pm \tanh \left( x\right)}.
\end{equation}
This is because the potential $V$ now has two degenerate vacua: the trivial one at $|\phi|=0$ and the broken one at $|\phi|=\sqrt{2}$. Obviously, the static kink can be multiplied by an arbitrary constant phase factor. The energy of the kink is 1 and carries a zero amount of $U(1)$ charge. 

For $\beta \in \left(\frac{1}{4},\frac{1}{3}\right)$ the broken-phase vacuum transforms into a local minimum. Thus, the kink separates the true and the false vacua, and cannot any longer be a static solution. In contrast, it starts to accelerate due to the static force triggered by the potential difference between the true and false vacua. 

Furthermore, it is known that the false vacuum of the potential $V$ can support a sphaleron (critical bubble), that is, a static unstable (saddle point) solution. In fact, there is a constant phase sphaleron that interpolates between the false vacuum $\phi=\phi_\textrm{min}$ (at $x\to \pm \infty$) and $\phi=\phi_0$ (at $x=0$). The value of the field at the center of the sphaleron, $\phi_0$, is set by the condition $V(|\phi_0|)=V(|\phi_\textrm{min}|)$. Depending on the perturbation, the false vacuum sphaleron decays into the true vacuum or into a false-vacuum oscillon for which the field quasi-periodically moves around the false vacuum. 

For $\omega \in (0,\omega_{min})$, there is a stationary version of the sphaleron that we call {\it $Q$-sphaleron} (in the literature it was also referred to as $Q$-hole or $Q$-bulge \cite{Nugaev:2016wyt}). Now, the profile interpolates between the local minimum $\phi_\textrm{min}^{\omega_0}$ of $V_{\omega_0}$ (at $x \to \pm \infty$) and $\phi=\phi_1$ at $x=0$. Again, $V(\phi_1)=V(\phi_\textrm{min}^{\omega_0})$. Similarly as in the case of the $Q$-kink, the $Q$-sphaleron has infinite energy if computed w.r.t. the true vacuum. Moreover, it carries an infinite amount of $U(1)$ charge. 

%%%%%%%%%%%%%%%%%%%%%%%%%%%%%%
\section{The bubble}
%%%%%%%%%%%%%%%%%%%%%%%%%%%%%%
%%%%%%%%%%%%%%%%%%%%%%%%%%%%%%
\subsection{Kink-antikink configuration}
\begin{figure}%[h!]
    \centering
    \includegraphics[width=1\textwidth]{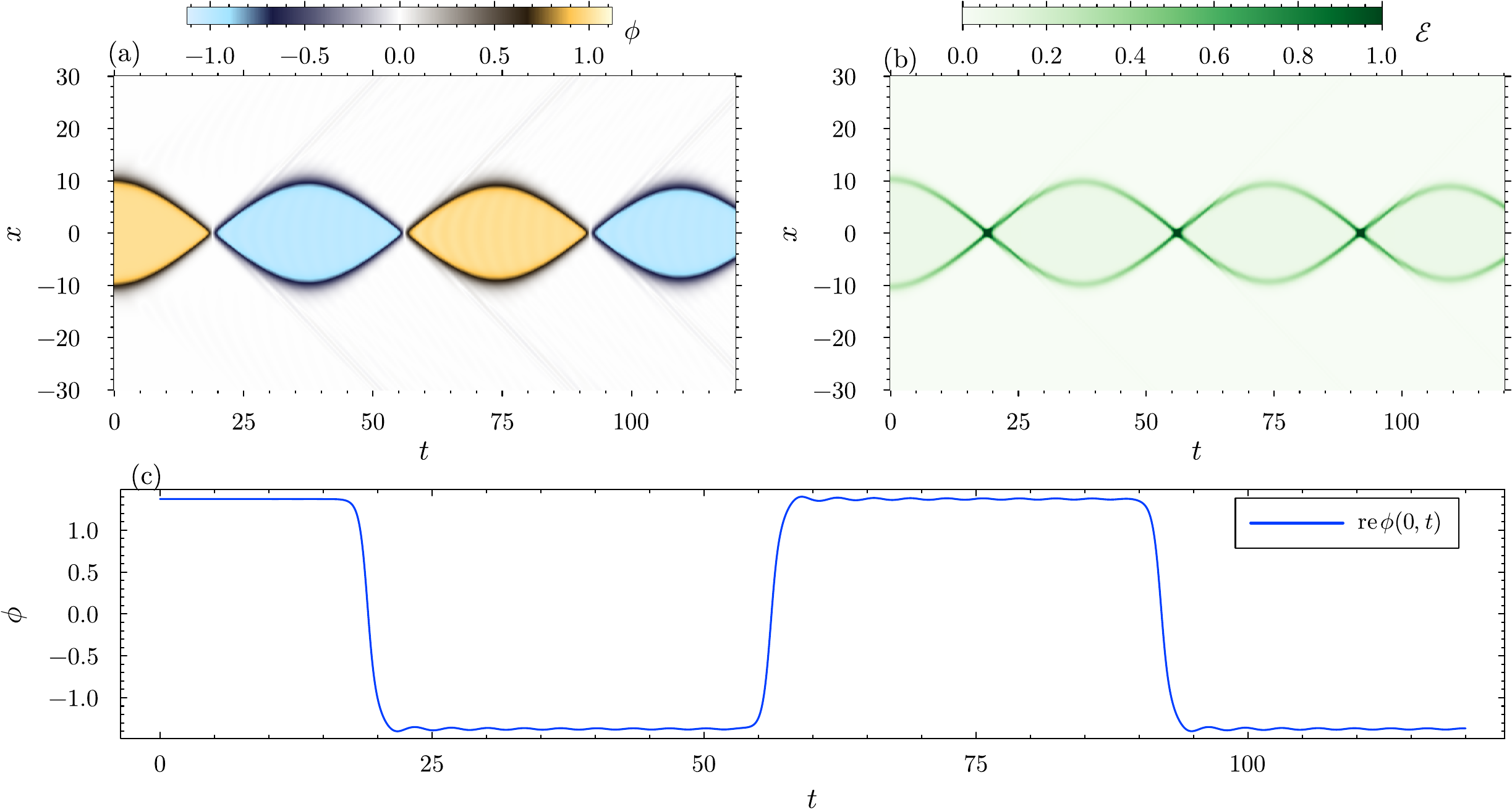}
    \caption{Evolution of the bubble for $a=10$. The first collapse is at $t=19.13$}
    \label{fig:bubble_pure}  
\end{figure}

%%%%%%%%%%%%%%%%%%%%%%%%%%%%%%
Finally, in the regime where the potential possesses a local minimum, there is a configuration in which the false (broken) vacuum is trapped inside the true (unbroken) vacuum. We call it a {\it bubble}
\begin{equation}
    \phi_B(x;a) = |\phi_\textrm{min}|\left(
        \sqrt{\frac{1+\tanh(x+a)}{2}} + \sqrt{\frac{1-\tanh(x-a)}{2}}
        \right)-|\phi_\textrm{min}|.
\end{equation}
It is formed by the static kink and antikink located at $x=\mp a$ respectively, each of them interpolating between the true and the false minimum. This configuration is not even an unstable static solution. The kinks immediately begin to attract each other. The force is proportional to the value of the potential in the false vacuum. The attraction leads to a sequence of bounces with the kinks re-created after each collision. This is a type of oscillon, sometimes called the {\it bion}, now based on the true vacuum, see Fig. \ref{fig:bubble_pure} where we plot the time evolution of the real component of the field, the energy density and the value of the real component of the field at origin. In this plot, and in the rest of this section, we assume that $\beta=0.26$, which is close to the critical value $1/4$. In this case, the broken false vacuum is very well pronounced and is not too high above the true vacuum. Consequently, the attractive force between the kinks, that is, the bubble boundaries, is relatively small. The resulting bion is a long-lived structure. 

In the remaining part of this section, we will show that the unstable bubble structures can be temporarily stabilized by small perturbations. 

%%%%%%%%%%%%%%%%%%%%%%%%%%%%%%
\subsection{The false vacuum and the Goldstone modes}
%%%%%%%%%%%%%%%%%%%%%%%%%%%%%%
The existence of a symmetry-breaking vacuum (false or true as in the case of the Peccei-Quinn $Q$-balls \cite{Kawasaki:2025nsi, Kobayashi:2025qao}) has some important consequences on the dynamics of the complex field and, in particular, on the dynamics of the $Q$-balls. This originates in the fact that the broken-symmetry vacuum supports a massless Goldstone (phase) mode that, at linear order, decouples from a massive (amplitude) mode. 

To see this let us consider a small perturbation of a given static or stationary solution $\Phi(x,t)=f(x)e^{i\omega_0 t}$,
\begin{equation}
    \phi(x, t) = f(x)e^{i\omega_0 t} + A\xi(x,t), 
\end{equation}
where the real parameter $A$ is the amplitude of the small perturbation. We plug it into the equation of motion and keep the linear terms in the amplitude
\begin{equation}
%\ddot \xi - \xi''+\left[V'(f_K^2)+V''(f_K^2)f_K^2\right]\xi + V''(f_K^2)f_K^2\xi^*=0 \,.
\ddot \xi - \xi''+\left[V'(f^2)+V''(f^2)f^2\right]\xi + V''(f^2)f^2\xi^*=0 \,.
\label{pert}
\end{equation}
For oscillating modes, it is consistent to consider perturbations $\xi$ of the form \cite{Ciurla:2024ksm}
\begin{equation}\label{decomposition}
    \xi(x,t) = e^{i(\omega_0+\rho) t}\eta_1(x) + e^{i(\omega_0-\rho) t}\eta_2(x) \, .
\end{equation}
Thus we arrive at a coupled two-channel problem with two components $\eta_1$ and $\eta_2^*$
\begin{equation}\label{linear}
    L\begin{bmatrix}
        \eta_1\\\eta_2^*
    \end{bmatrix}
= 
    \begin{bmatrix}
    (\omega_0+\rho)^2& 0\\
    0&(\omega_0-\rho)^2
\end{bmatrix}
\begin{bmatrix}
        \eta_1\\\eta_2^*
    \end{bmatrix},
\end{equation}
where
\begin{equation}
L=
\begin{bmatrix}
    D& S\\
    S&D
\end{bmatrix}
\;\;\; \mbox{and} \;\;\;   D=-\frac{d^2}{dx^2}+U + S. 
\end{equation}
The potentials entering the spectral problem are
\begin{equation}\label{potentialU}
    U = V'(f^2)=1-2f^2+3\beta f^4
\end{equation}
\begin{equation}\label{potentialS}
    S = f^2V''(f^2)=f^2(-2+6\beta f^2).
\end{equation}

Let us now consider a static solution for which $\omega_0=0$. The spectral problem reduces to
\begin{equation}\label{linear}
    \left(L - \rho^2\right)\begin{bmatrix}
        \eta_1\\\eta_2^*
    \end{bmatrix}
= 0,
\end{equation}
which can be written in a decoupled form  
\begin{equation}\label{linear-new}
    \begin{bmatrix}
    -\frac{d^2}{dx^2}+U_G& 0\\
    0&-\frac{d^2}{dx^2}+U_A
\end{bmatrix}
\begin{bmatrix}
        \tilde{\eta}_1 \\\tilde{\eta}_2
    \end{bmatrix}=\rho^2
\begin{bmatrix}
        \tilde{\eta}_1 \\\tilde{\eta}_2
    \end{bmatrix},
\end{equation}
where $\tilde{\eta}_1=\eta_1-\eta_2^*$ and $\tilde{\eta}_2=\eta_1+\eta_2^*$ are the imaginary and real perturbations (modes) respectively,
\begin{align}
\phi(x,t)&=f(x)e^{i\omega_0t}+Ae^{i\omega_0t}\left[\text{Re}\left(\tilde\eta_2e^{i\rho t}\right)+i\text{Im}\left(\tilde\eta_1e^{i\rho t}\right)\right].
\end{align}
We take these modes to have unit $L^2$ norm.  In addition, 
\begin{equation}
    U_G=U, \;\;\;\; U_A=U+2S
\end{equation}
are the new potentials, which correspond to the Goldstone and amplitude excitations.

In the true vacuum, where $f=0$, the potentials reduce to $U(x)=1$ and $S(x)=0$. In this limit, the decoupled equations for $\tilde{\eta}_1$ and $\tilde{\eta}_2$ describe small perturbations with the mass parameter $m_0=1$. Hence, both perturbations are massive. 

In the broken vacuum, where $f=|\phi_\textrm{min}|\neq 0$, we get 
\begin{equation}
    U_f=0\,,\qquad S_f=\frac{2}{3\beta}\left(1-3\beta+\sqrt{1-3\beta}\right),
\end{equation}
where $U_f=U(f=|\phi_{min}|)$, and $S_f=S(f=|\phi_{min}|)$.
Clearly, $\tilde{\eta}_1$ is a massless perturbation, $m_1=0$ that is a Goldstone mode, while $\tilde{\eta}_2$ is still a massive perturbation with $m_2^2=2S_f$. This mass varies from $m_2=2$ for $\beta=\frac{1}{4}$ to zero for $\beta=\frac{1}{3}$, where the false vacuum ceases to exist. Note that for $\beta \in \left(0.32, 1/3 \right)$, $m_2<m_0$ and even this mode is lighter than the mass threshold. Finally, we underline that at linear order these modes do not mix with each other.
   
\begin{figure}%[h!]
    \centering
    \includegraphics[width=1.0\textwidth]{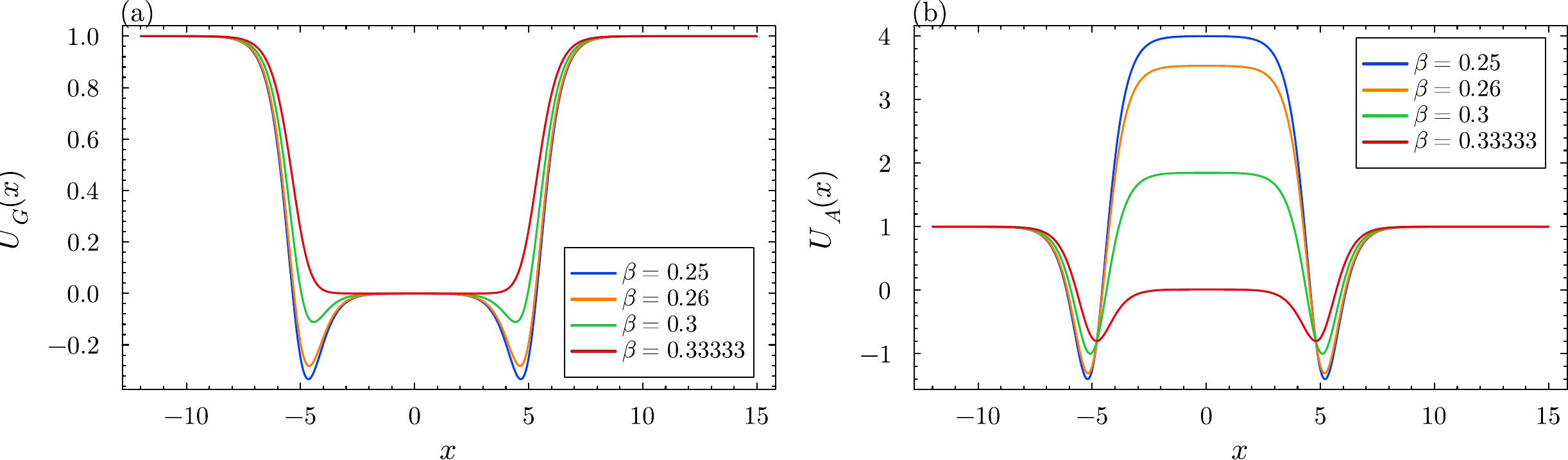}
    \caption{Goldstone and amplitude potentials generated by the bubble for different values of $\beta$ and for $a=5$.}
    \label{fig:bubble_potentials}  
\end{figure}

Now, we move on to the bubble $\phi_B(x)$, which describes the false vacuum trapped inside the true vacuum, and perform a linear perturbation around a fixed bubble configuration. Of course, there is always an unstable, negative energy mode, as the bubble is not a critical point of the action. Note that this bubble is a time-independent configuration (however, it is not a static solution) and the spectral problem can be written in the diagonal form (\ref{linear}) but with different potentials $U_G$ and $U_A$. We plot them in Fig. \ref{fig:bubble_potentials}.

The qualitative picture is clear. The purely imaginary perturbation is the massless Goldstone mode inside the bubble, while outside the bubble it has a nonzero mass. Hence, once excited inside the bubble, the imaginary perturbation exerts a positive radiation pressure on the boundary of the bubble. If the imaginary mode is excited outside, it leads to negative radiation pressure. Both scenarios can stabilize the bubble against collapse. 

The real excitation, which always corresponds to a massive, amplitude mode, has two different regimes. For $\beta < 0.32$, its mass is larger in the broken vacuum inside the bubble. Thus, if excited inside the bubble, it leads to a negative radiation pressure, which brings the boundaries of the bubble together. This acts like an additional attractive force that facilitates the collapse of the bubble. On the other hand, for $0.32 < \beta < \frac{1}{3}$, the amplitude mode has a smaller mass in the false vacuum. Therefore, it helps stabilize the bubble in the same way as the Goldstone mode does. 

\begin{figure}%[h!]
    \centering
    \includegraphics[width=1\textwidth]{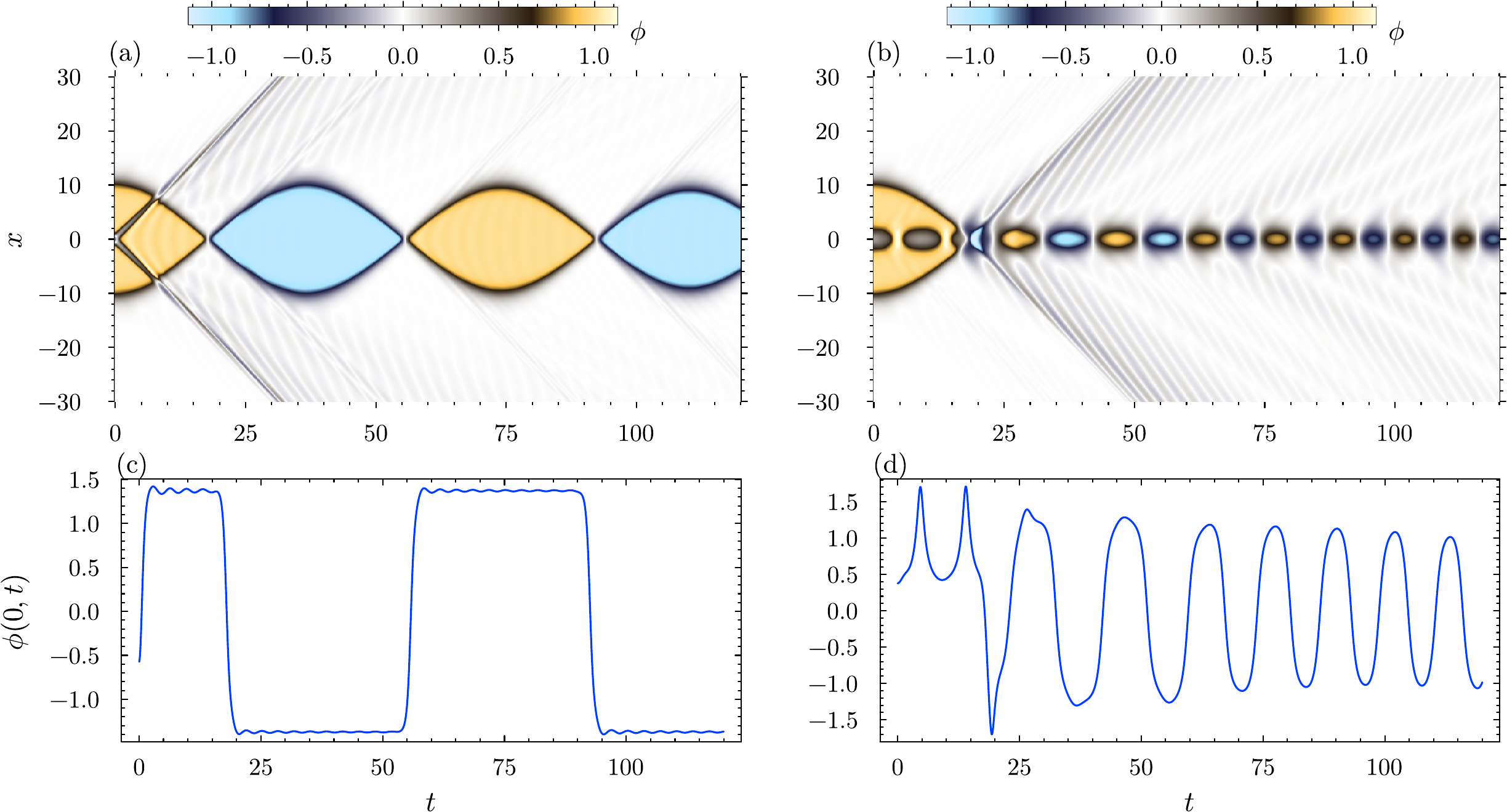}
    \caption{Evolution of the bubble for $\beta=0.26$ and $a=10$ with additional perturbation $ A\exp(\alpha x^2)$, see further description in the text. The first collapse is at $t=17.97$ (a,c) and at $t=17.84$ (b,d).}
    \label{fig:bubble_real}  
\end{figure}
 
\begin{figure}%[h!]
    \centering
    \includegraphics[width=0.95\textwidth]{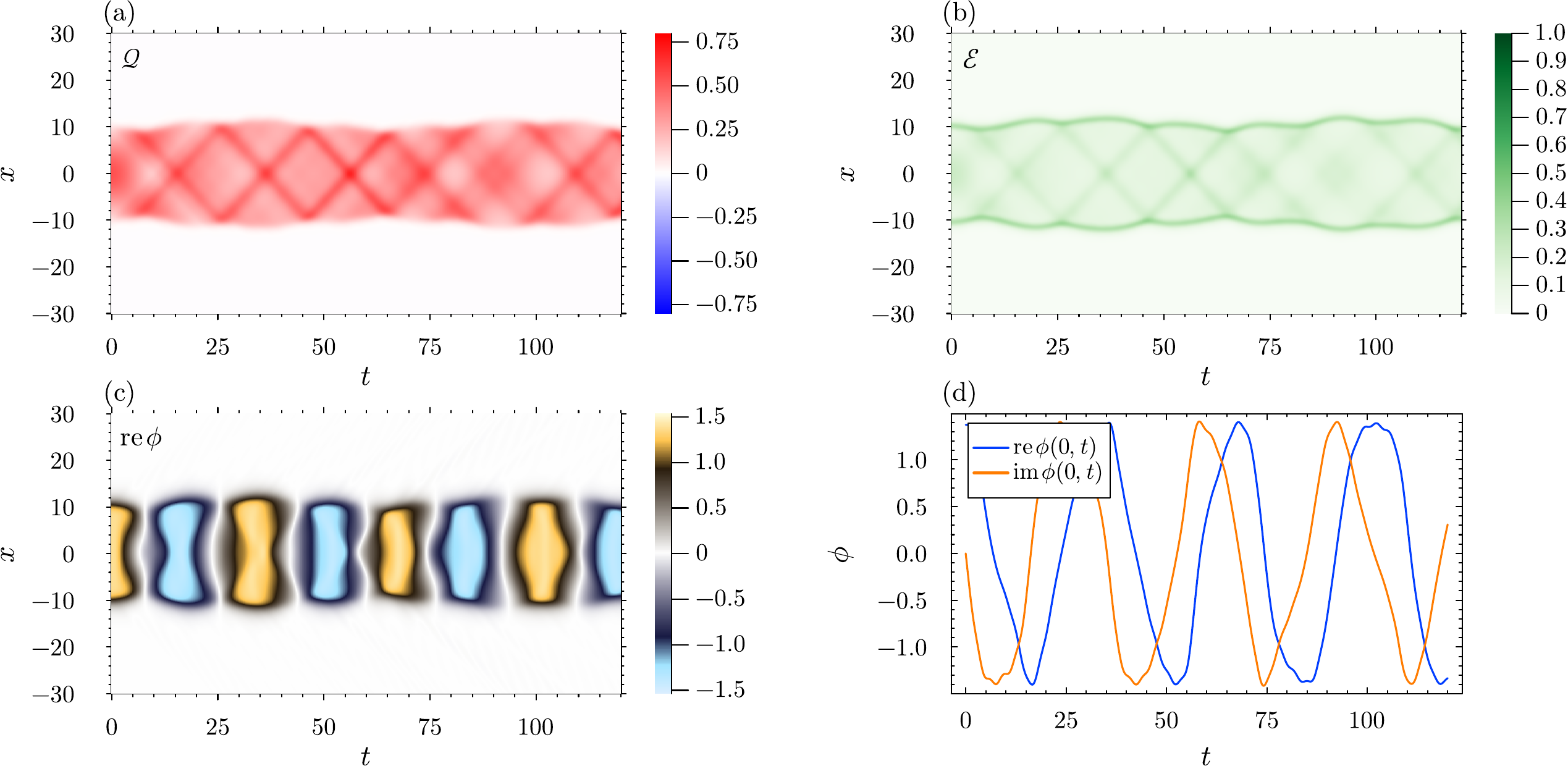}
    \caption{Stabilization of a bubble for $a=10$ with initial perturbation $ -i\exp(-(x/8)^2)\sin(0.4t)$. Here $\beta=0.26$.}
    \label{fig:bubble_Goldstone}  
\end{figure}

In Fig. \ref{fig:bubble_real} we show the evolution of the bubble for $\beta=0.26$ and $a=10$ while a real perturbation, $Ae^{\alpha x^2}$, is added. In the left panels, where $A=-2$ and $\alpha=1.1$, the bion survives the perturbation. In the right panels, where $A=-1$ and $\alpha=0.5$, the perturbation creates an oscillon in the false vacuum that collides with the boundaries of the collapsing bubble and eventually forms an oscillon in the true vacuum. Here, the bubble (or bion) completely annihilates. As expected from the previous analysis, in both cases the first collapse occurs earlier than in the evolution of the pure bubble, Fig. \ref{fig:bubble_pure}. Specifically, it occurs at $t=17.97$ and $t=17.84$.  

An imaginary excitation acts differently. Namely, it stabilizes the bubble, see Fig. \ref{fig:bubble_Goldstone}. The perturbation clearly generates the Goldstone mode, a wave traveling at the speed of light, which bounces from the boundaries of the bubble. Now, the bubble does not collapse for quite a long time. Note also that the imaginary perturbation adds a non-zero amount of the $U(1)$ charge. Hence, in this particular case, the resulting solution can be viewed as a perturbed thin-wall $Q$-ball. However, the bubble can also be stabilized by Goldstone modes, which carry the zero total charge.

%%%%%%%%%%%%%%%%%%%%%%%%%%%%%%
\subsection{Trapped Goldstone modes and spectral walls}
%%%%%%%%%%%%%%%%%%%%%%%%%%%%%%

The stabilization of the bubble via some perturbations can also be understood as the excitation of bound modes hosted in the potential wells $U_G$ and $U_A$, generated by the bubble solution, see Fig. \ref{fig:bubble_potentials}. These bound modes are simply the Goldstone modes of the false vacuum trapped inside the bubble. Since they are trapped in the cavity, there is a finite number of discrete bound modes, standing waves, $\tilde{\eta}_{k}(a)$, whose frequencies $\omega_k$ and even number depend on the size of the potential well, that is, the size of the bubble $2a$. 

\begin{figure}%[h!]
    \centering
    \includegraphics[width=0.85\textwidth]{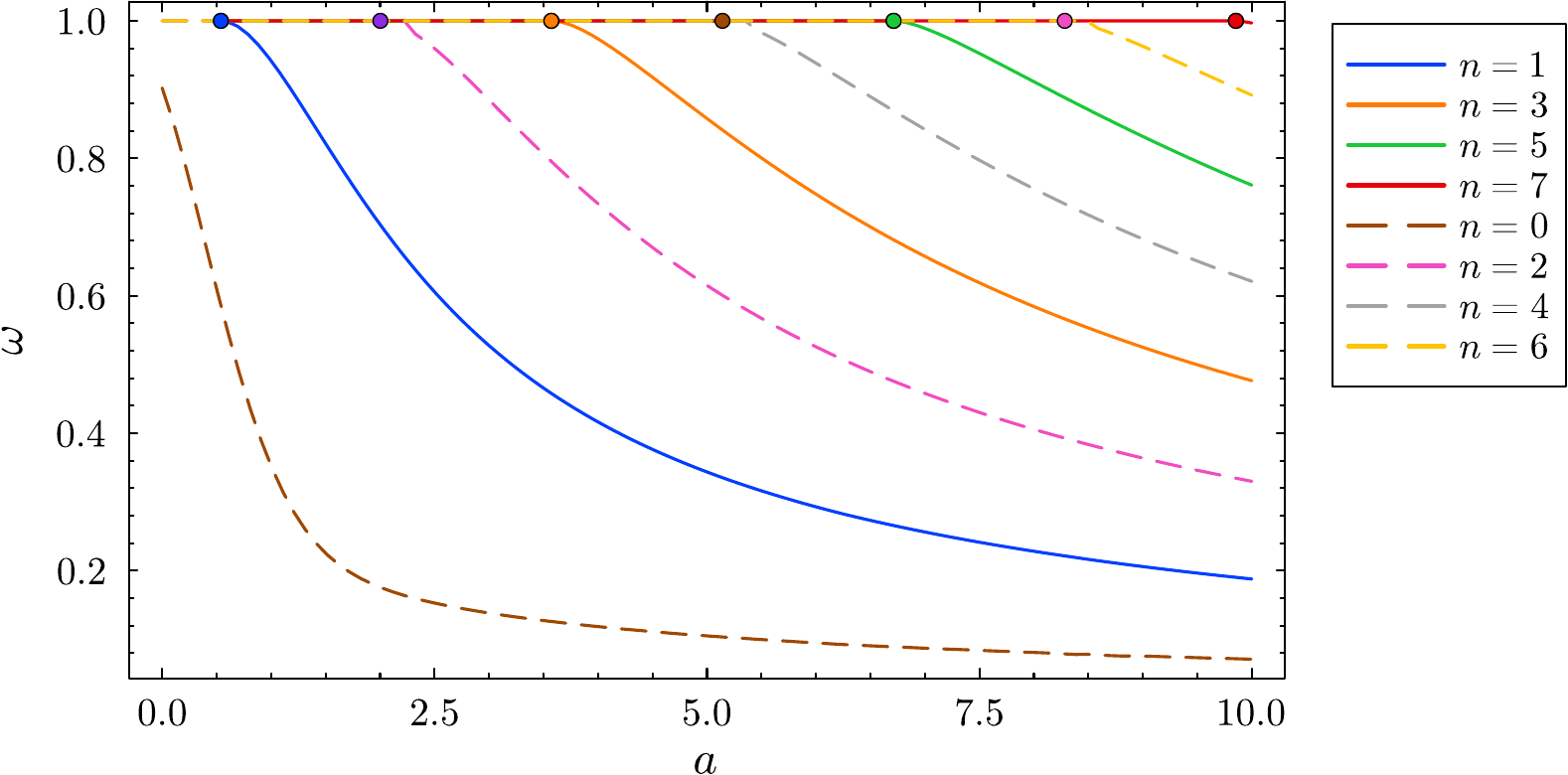}
    \caption{The flow of the spectral structure of the bubble with its size. We plot the odd and even trapped Goldstone's modes. Here, $\beta=0.26$.}
    \label{fig:spectral structure}  
\end{figure}

In Fig. \ref{fig:spectral structure} we show how the frequencies of such trapped Goldstone modes vary with the size of the bubble, for $\beta=0.26$. As the size increases, the frequencies decrease and one by one new modes appear. This has two important consequences.

%\vspace*{0.2cm}

First of all, it leads to the existence of a new mode generated {\it repulsive force} between the kinks forming the boundaries of the bubble. This comes from the fact that in the leading order, quadratic, approximation each mode contributes to the potential energy of the solution as 
\begin{equation}
    E_{mode} = \frac{1}{2} A^2 \omega^2(a),
\end{equation}
where we assume that the mode has unit $L^2$ norm. Here, we have two degrees of freedom: the amplitude $A$ of a bubble-size-dependent oscillator and the size $a$ of the bubble itself. Both variables vary in time and, at leading order, constitute the following effective Lagrangian 
\begin{equation}
    L[A,a] = \frac{1}{2} M(a)\dot{a}^2 + \frac{1}{2}\dot{A}^2- \frac{1}{2} A^2 \omega^2(a) - V(a). 
\end{equation}
Here, 
\begin{equation}
  M(a)=\int_{-\infty}^\infty \left( \frac{d \phi_B(x;a)}{dx} \right)^2 dx  
\end{equation}
is the mass (energy) of the bubble and 
\begin{equation}
  V(a)=\int_{-\infty}^\infty \left[ \left( \frac{d \phi_B(x;a)}{dx} \right)^2 + \phi^2_B(x;a) -\phi^4_B(x;a)+\beta \phi^6_B(x;a) \right]dx  
\end{equation}
 is a contribution to the potential energy from the attraction between the boundaries of the bubble. $V(a)$ has a single zero at $a=0$ and increases monotonically with $a$.

In the adiabatic approximation, where the velocity of the bubble boundaries is small, $\dot{a} \sim O(\epsilon)$, the mode amplitude evolves as \cite{AlonsoIzquierdo:2024nbn}
\begin{equation}
    A(t)=\frac{C}{\sqrt{\omega(a(t))}} \cos \left( \int_0^t \omega (a(t')) dt'\right),
\end{equation}
where $C$ is an adiabatic invariant that remains constant despite the fact that the frequency of the oscillator changes with the size of the bubble. Averaging $A^2$ over the period gives $\langle A^2 \rangle=\frac{1}{2} \frac{C^2}{\omega}$. This leads to a reduced Lagrangian
\begin{equation}
    L[A,a] = \frac{1}{2}M(a)\dot{a}^2 - \frac{1}{2}  C^2 \omega (a) - V(a). 
\end{equation}
Thus, the oscillator dynamics generates a potential energy that is linear in the frequency
\begin{equation}
    E_{osc} = \frac{1}{2}  C^2 \omega (a).
\end{equation}
As $\omega$ decreases with $a$, it is energetically preferred to increase the size of the bubble. Thus, the repulsive force appears. This can stabilize the bubble. Of course, if frequencies of the modes grow with the size of the bubble, then they contribute to an attractive force.  

\begin{figure}%[h!]
    \centering
    \includegraphics[width=1.0\textwidth]{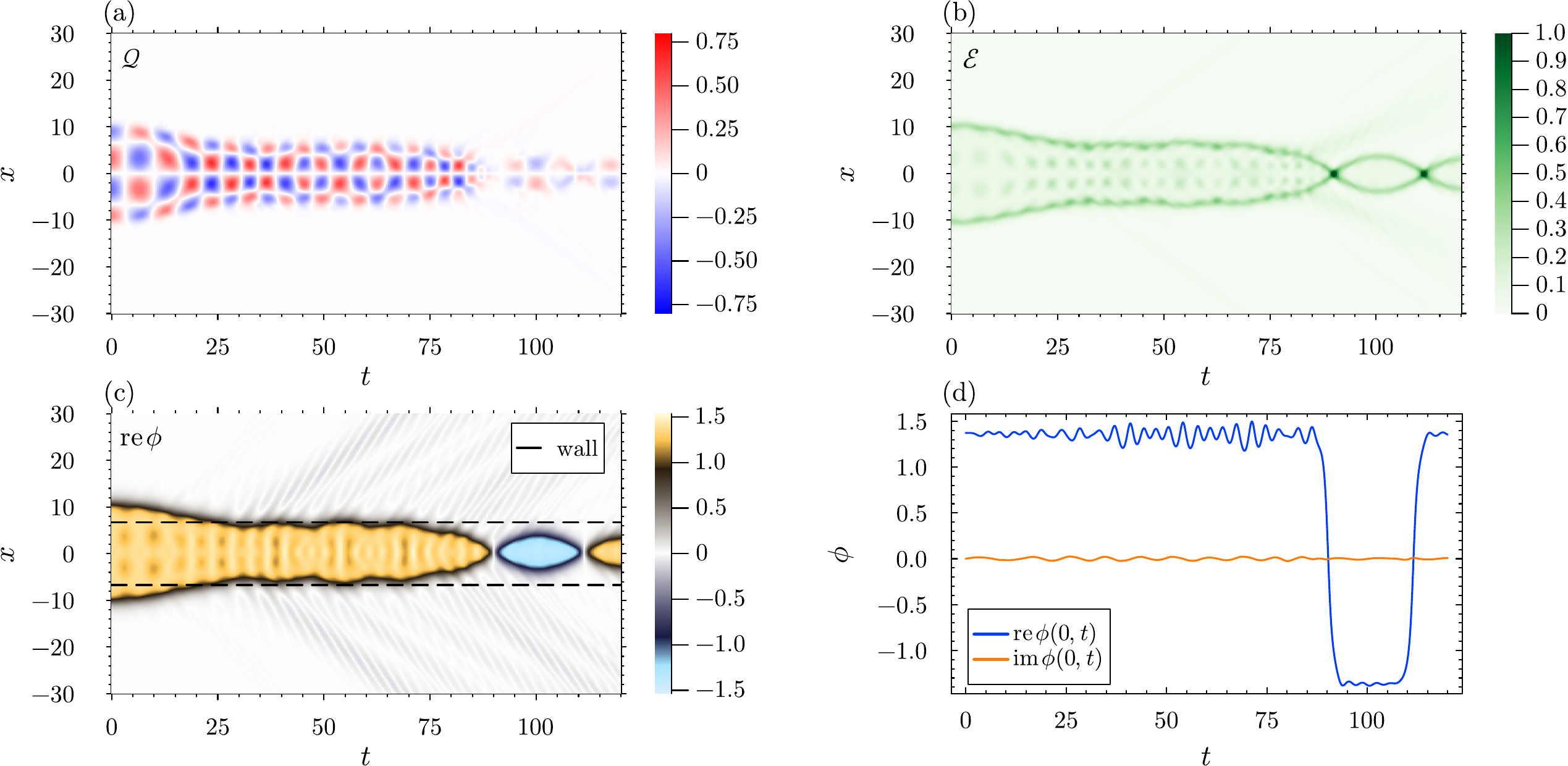}
    \caption{Spectral wall in the bubble dynamics for $\beta=0.26$. Initially, the bubble with $a=10$ has the third trapped mode excited with the amplitude $A=2.72$. The dashed line denotes the position of the corresponding spectral wall.}
    \label{fig:SW}  
\end{figure}

Secondly, as the size of the bubble decreases, the trapped Goldstone modes hit the mass threshold one by one. Such a transition of a bound mode to the continuum spectrum triggers {\it the spectral wall phenomenon} \cite{Adam:2019xuc}. If such a mode, $\tilde{\eta}_k$, is excited, then the bubble boundaries experience an additional repulsive force once the bubble size is reduced to the critical value, $2a_{sw}^{(k)}$, at which the mode hits the mass threshold. At this point, the mode formally becomes a non-normalizable threshold mode and apparently the soliton tries to avoid this possibility. Thus, exactly for this size, the bubble can form a quite stable, long-lived state, provided the mode is excited with a fine-tuned amplitude. In addition, passing through a spectral wall leads to rather significant radiation bursts. For the example shown in Fig. \ref{fig:spectral structure} the positions of the spectral walls are: $a_{sw}^{(1)}=0.5392, \, a_{sw}^{(2)}=2.0026, \, a_{sw}^{(3)}=3.5714, \, a_{sw}^{(4)}=5.1421, \, a_{sw}^{(5)}=6.7128, \, a_{sw}^{(6)}= 8.2834,  \, a_{sw}^{(7)}=9.8541$.

We can see the effect of spectral walls clearly in our numerics. In Fig. \ref{fig:SW} we present the evolution of the bubble with the third mode (second odd mode) excited. Initially, the amplitude of this mode is $A=2.72$ and its frequency is $\omega_2=0.48007$ as the half-separation is $a(t=0)=10$. At the initial stage of the evolution, the attractive force between the bubble boundaries wins, and the bubble shrinks. However, as $a$ decreases, the frequency of the mode approaches the mass threshold. At this stage, a relatively long-lived state is formed where the bubble boundaries keep (approximately) at a fixed distance, which agrees very well with the critical value $a_{sw}^{(3)}$. Eventually, as the amplitude of the mode decreases due to radiation, the configuration passes through the spectral wall. Then, the bubble collapses to the bion. In the plot, we show the evolution of the charge density, energy density, real component of the field, and real and imaginary components at the origin. 

We found analogous spectral walls for the lower and higher modes. As always, this is a selective phenomenon, which means that the system sees a particular spectral wall only if the corresponding mode is sufficiently excited. It remains blind to all other spectral walls. For lower modes, due to the rather strong attraction of the bubble boundaries, one needs a higher amplitude, which, via nonlinearities of the model, leads to the excitation of other modes. This makes the spectral wall picture less transparent. 

%%%%%%%%%%%%%%%%%%%%%%%%%%%%%%
\section{$QQ^*$ collisions}
%%%%%%%%%%%%%%%%%%%%%%%%%%%%%%
Here we will consider the collision of $QQ^*$. Thus, we assume that initially the well-separated $Q$-balls are boosted toward each other with initial velocity $v$
\begin{equation}
 \phi(x,t)=f_{\omega_0}(\gamma(x+x_0-vt))e^{i\omega_0 \gamma(t-vx)+i\delta} + f_{\omega_0}(\gamma(x-x_0+vt))e^{-i\omega_0 \gamma(t+vx)},     
\end{equation}
where $\gamma=1/\sqrt{1-v^2}$ and $\delta$ are the initial phase shift of one of the $Q$-ball. The initial separation $x_0$ is assumed to be large enough to treat the initial objects as independent. 

In the previous sections, we have described the possible states that may participate in collisions of the $Q$-balls. First of all, we have the stable solutions that are the $Q$-balls themselves. They are true final states in the scatterings. They host bound modes that may trigger the resonant energy transfer mechanism playing a crucial role in collisions of topological solitons such as kinks and vortices. In a scattering process, two solitons can temporarily transfer their kinetic energy into internal degrees of freedom, i.e. the internal modes (bound modes or quasinormal modes). If there is an attractive static intersoliton force, as in the case of a kink-antikink pair,  or a mode generated force, as in the case of the BPS vortices, after collision the solitons can have too little kinetic energy to overcome the attraction. Hence, they collide again. This can lead to a complete annihilation of the solitons, usually via the formation of a long-lived oscillon (bion). Sometimes, during the collision, a sufficient amount of energy is returned to the kinetic motion for the solitons to eventually escape to infinity as the final state. This mechanism amounts to a complicated and chaotic pattern in the final state formation in 1+1 \cite{Sugiyama:1979mi, Campbell:1983xu, Manton:2021ipk} and 2+1 dimensions \cite{AlonsoIzquierdo:2024nbn, Krusch:2024vuy, Maxi}, where the annihilation regions, that is, the {\it bion chimneys}, are separated by the {\it bounce windows} with the solitons recreated in the final states. 

\begin{figure}%[h!]
 \includegraphics[width=1.05\textwidth]{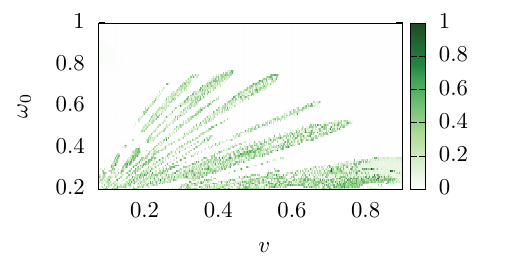}
 \includegraphics[width=1.05\textwidth]{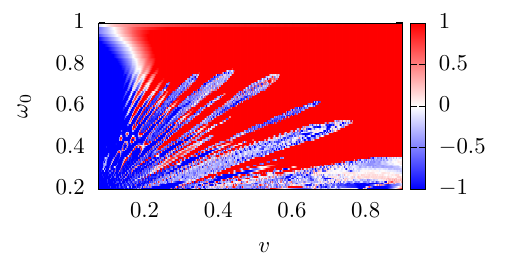}
 \caption{Energy density at the origin (upper) and total charge at the semi-line $x>0$ (lower) for the $QQ^*$ scattering as a function of the initial velocity $v$ and internal frequency $\omega_0$ of the scattered $Q$-balls. The values are computed at the final time of the evolution $t=200$. Here $\beta=0.26$.}
    \label{fig:scans} 
\end{figure}

Importantly, in addition to the internal modes, there is another factor that has a great impact on the collisions of the $Q$-balls. This is the existence of the ephemeral states, i.e., the polarized $Q$-balls and the bubble. Through perturbation of the colliding $Q$-balls one gets the whole family of polarized $Q$-balls, which includes not only $Q$-balls with the internal modes excited but also charged, as well as zero-charge, oscillons. Note that the $U(1)$ charge can be smoothly deformed by such perturbations. This is an important modification, if compared with the case of topological solitons, where the charge is quantized.  
Of course, the polarized $Q$-balls (e.g., charged oscillons) are not ultimately stable but can be very long-lived states that slowly decay via radiation to other $Q$-balls (nonzero net charge) or the vacuum (zero net charge). In any case, polarized $Q$-balls can easily show up in collisions of the $Q$-balls as intermediate (excited) states.

Finally, there is also the bubble, i.e., a zero-charge unstable configuration that exists for the potentials with a false broken vacuum. Although short-lived, it can be very easily stabilized by excitation of the Goldstone modes. 

As we shall see in the rest of the paper, the internal modes and ephemeral states are the main ingredients governing the $QQ^*$ collisions.

\vspace*{0.2cm}

There are three obvious regimes that may lead to qualitatively different dynamics. They arise from two regimes of the potential and two regimes of the $Q$-balls. Namely, the potential can be with ($\beta<1/3$) or without ($\beta>1/3$) the false vacuum, which results in the existence or nonexistence of the bubble. The existence of a false vacuum will be especially important when the $Q$-ball probes the thin-wall regime, $\omega_0\to \omega_{min}$. In the thick-wall regime, $\omega_0\to 1$, the existence of the second vacuum is less important since the field does not probe such large field values.  Thus, in this limit, the particularities of the potential are expected to play a secondary role. Obviously, in the regime no false vacuum the complexity of the $QQ^*$ collisions can be exclusively related to internal modes. 

These regimes are clearly visible in Fig. \ref{fig:scans} where we show the energy density at the origin and the total $U(1)$ charge in the semiline $x>0$ in the final stage of evolution, $t_{f}=200$, where we considered a version of the model with the false vacuum $\beta=0.26$. In the thick-wall regime, where $\omega_0\to 1$, the collision has a fairly simple structure with a region of back-scattering (smaller velocities) and passing through (higher velocities). Interestingly, the $Q$-balls exist after the collisions, although in excited states. There is almost no annihilation. As expected, as we move towards the thin-wall regime, we find a chaotic behaviour. In particular, the characteristic spine structure of Fig. \ref{fig:scans} can be associated with the existence of surface substructure of the $Q$-ball, and bear striking similarities with the kink-antikink collisions in the Christ-Lee model \cite{Dorey:2023izf}. 
%%%%%%%%%%%%%%%%%%%%%%%%%%%%%%
\subsection{Thick-wall regime}
%%%%%%%%%%%%%%%%%%%%%%%%%%%%%%

\begin{figure}%[h!]
 \includegraphics[width=0.51\textwidth]{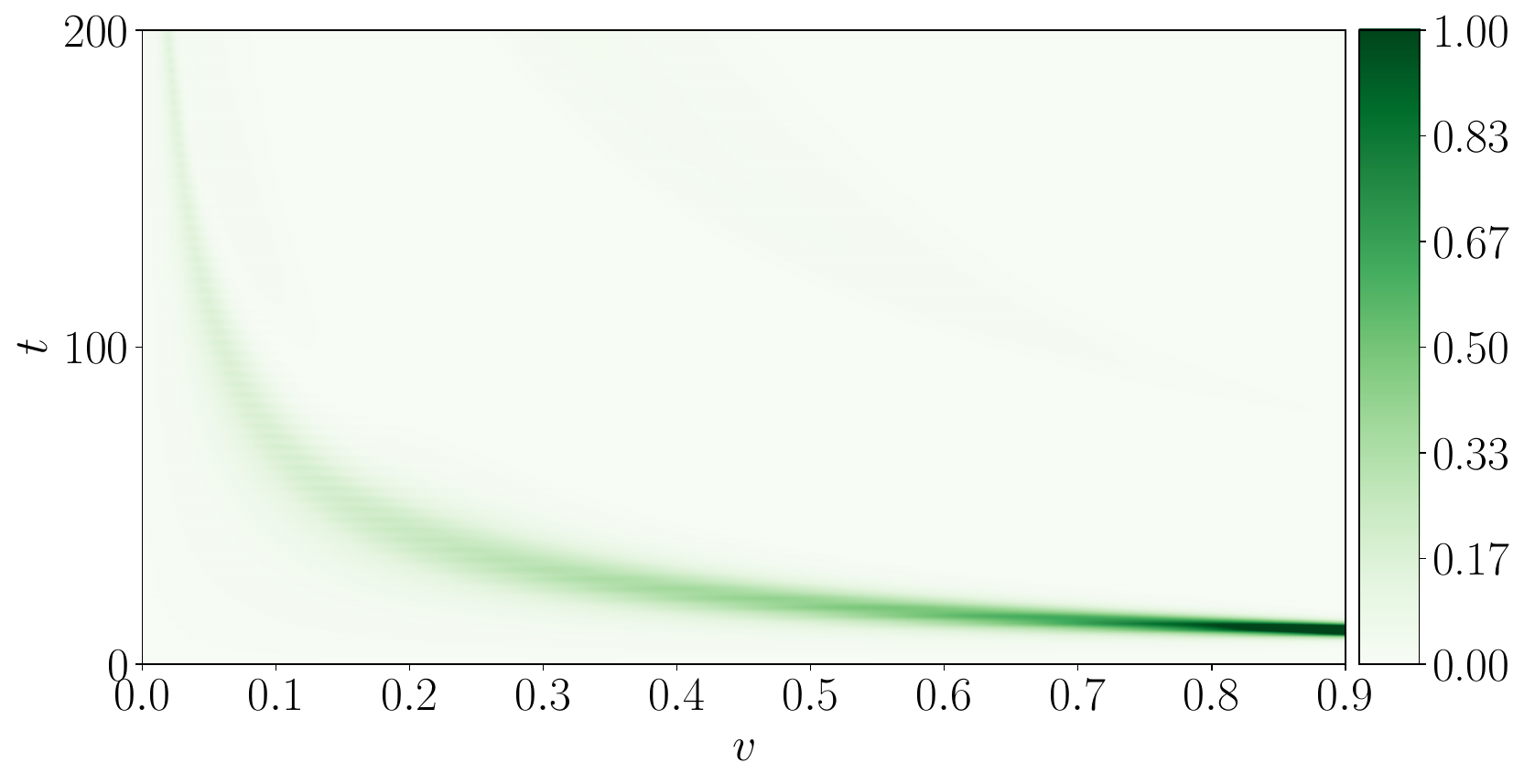}
 \includegraphics[width=0.51\textwidth]{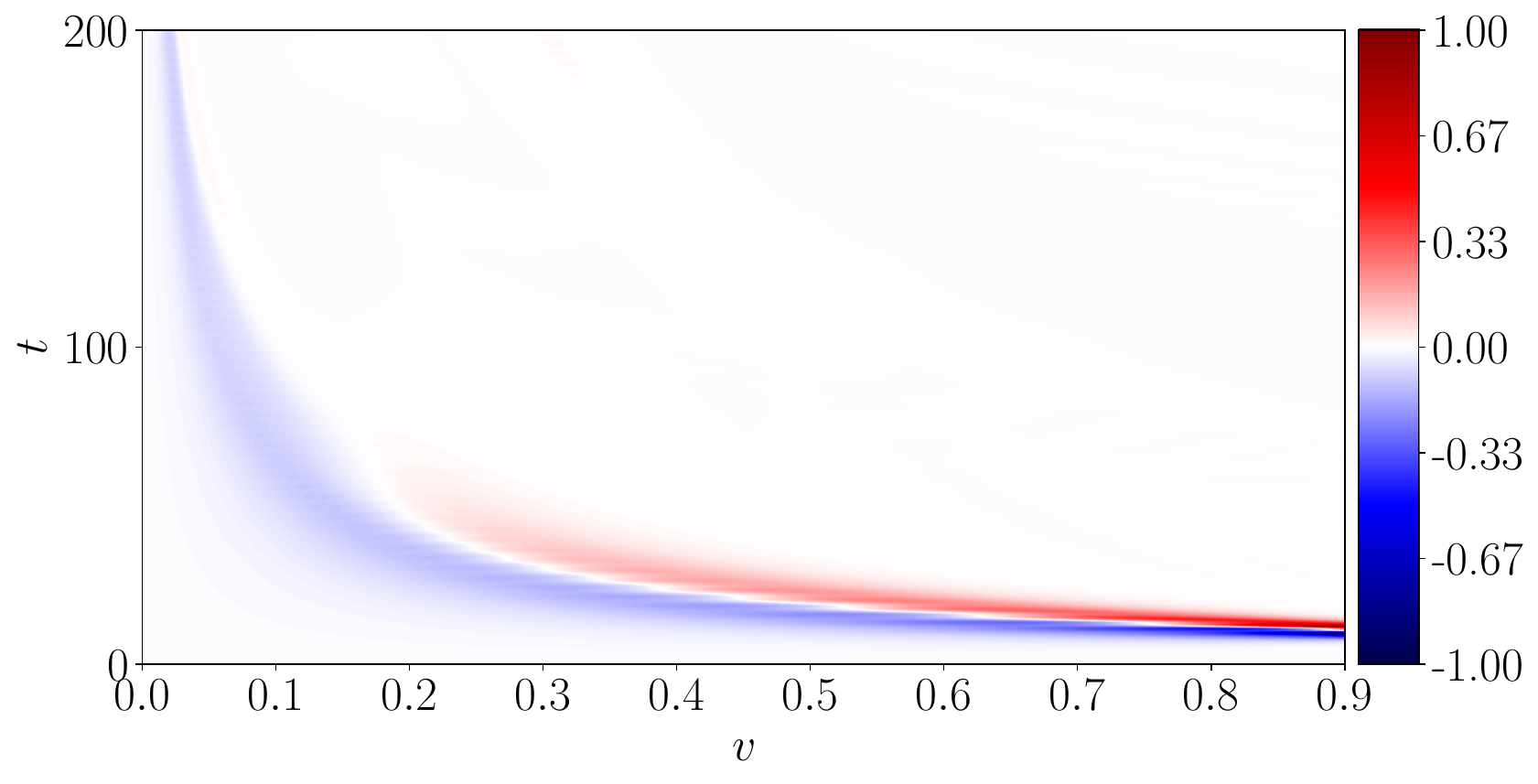}
 \caption{Evolution of the energy density $\mathcal{E}(x=0,t)$ and charge density at $x=0.05$ for different values of the initial velocity of the $Q$-ball. Here, $\beta=0.26$ and incoming $Q$-balls are in the {\it thick-wall regime}, $\omega_0=\pm 0.708$. Initial phase is $\delta=0$.}
    \label{fig:thick} 
\end{figure}

The $QQ^*$ collisions in the thick-wall limit have a very simple structure. Basically, there are only two scenarios: at smaller velocities the $Q$-balls backscatter, while at higher velocities they pass through each other. This does not depend on a particular form of the potential and existence of the additional vacuum. This is because in this limit the $Q$-balls are rather simple objects. They do not have an inner structure and do not host bound modes. Thus, resonant energy transfer is suppressed and no chaotic pattern in the collision is observed, see Fig. \ref{fig:thick}. This is consistent with the results known from scattering kinks, where the existence of internal modes (localized on each kink \cite{Sugiyama:1979mi, Campbell:1983xu, Manton:2021ipk} or on the two-soliton state \cite{Dorey:2011yw}) is typically a necessary condition for chaotic behaviour in the kink-antikink collisions.

%%%%%%%%%%%%%%%%%%%%%%%%%%%%%%
\subsection{Thin-wall regime without false vacuum}
%%%%%%%%%%%%%%%%%%%%%%%%%%%%%%
 
In the thin-wall regime $Q$-balls do possess vibrational modes that may trigger the resonant transfer mechanism. We begin with $\beta>1/3$, where there is no false vacuum, and the only ephemeral states are oscillons and other members of the family of the polarized $Q$-balls. Importantly, there is no bubble. 

\begin{figure}%[h!]
 \includegraphics[width=0.51\textwidth]{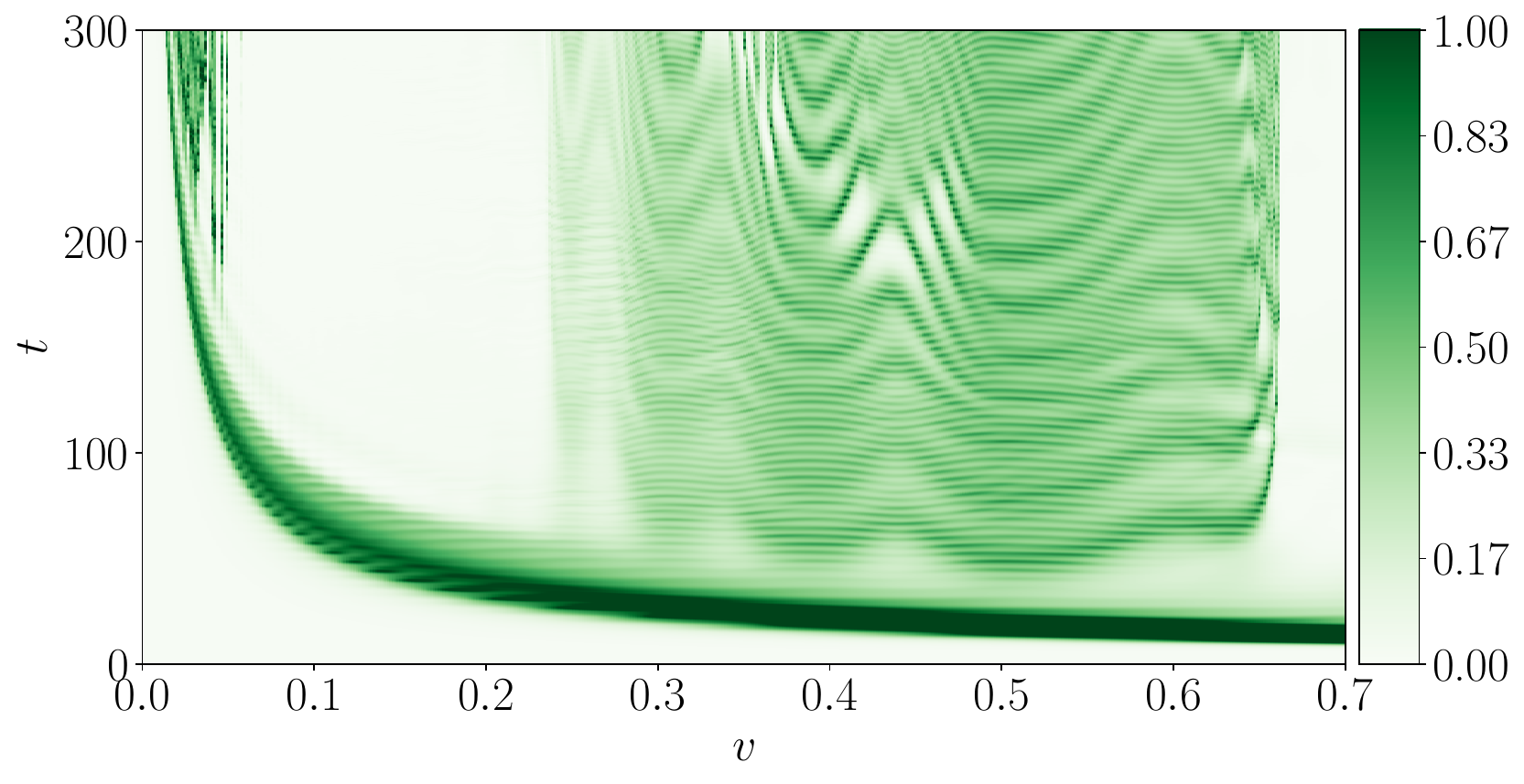}
 \includegraphics[width=0.51\textwidth]{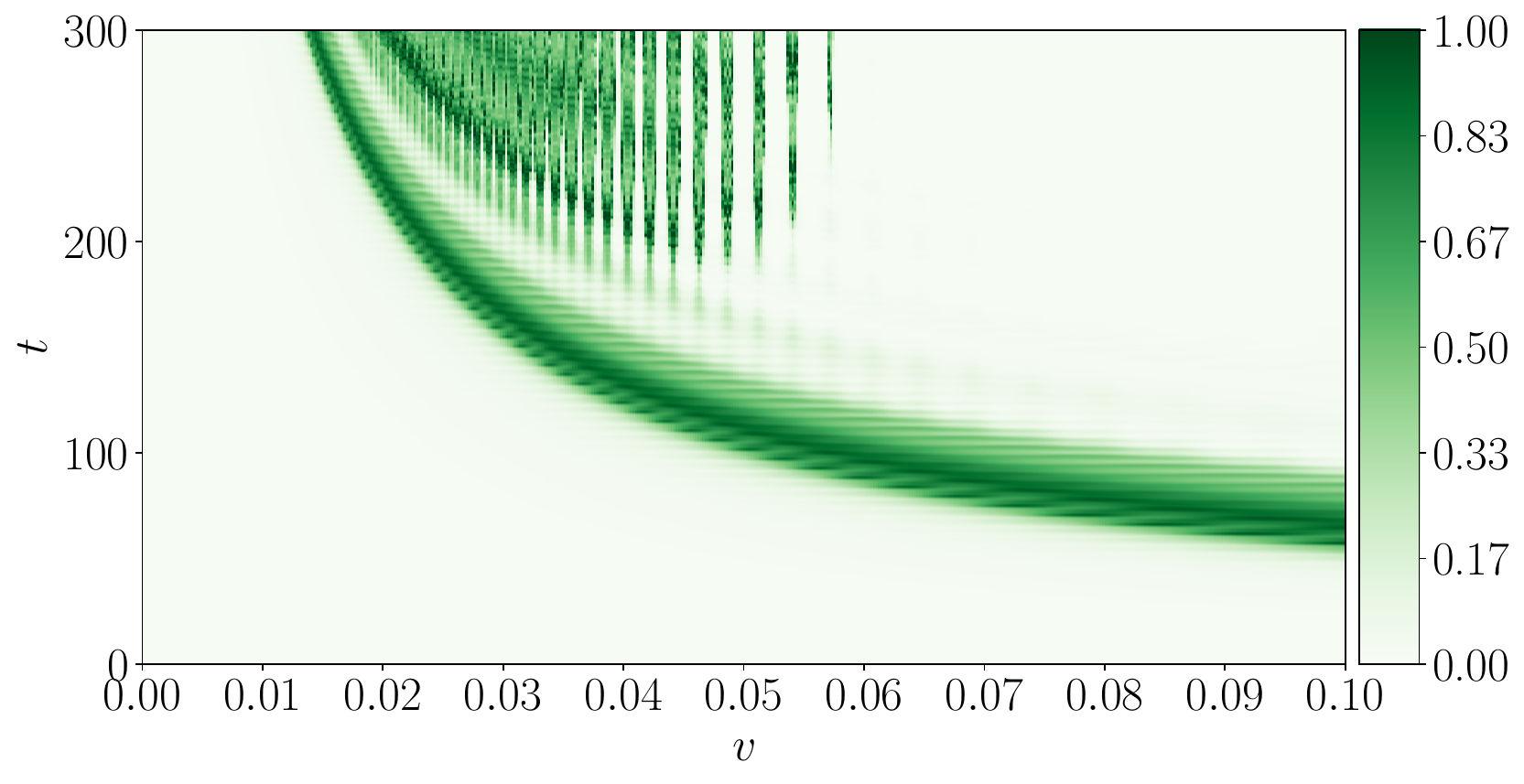}
 \caption{Evolution of the energy density $\mathcal{E}(x=0,t)$ in time for different values of the initial velocity of the $Q$-ball. Here, $\beta=0.5$ and incoming $Q$-balls are in the {\it thin-wall regime}, $\omega_0=\pm 0.708$. Initial phase is $\delta=0$. The right-hand plot shows a zoomed-in region of the left-hand plot.}
    \label{fig:no-bubble} 
\end{figure}

\begin{figure}%[h!]
 \includegraphics[width=0.50\textwidth]{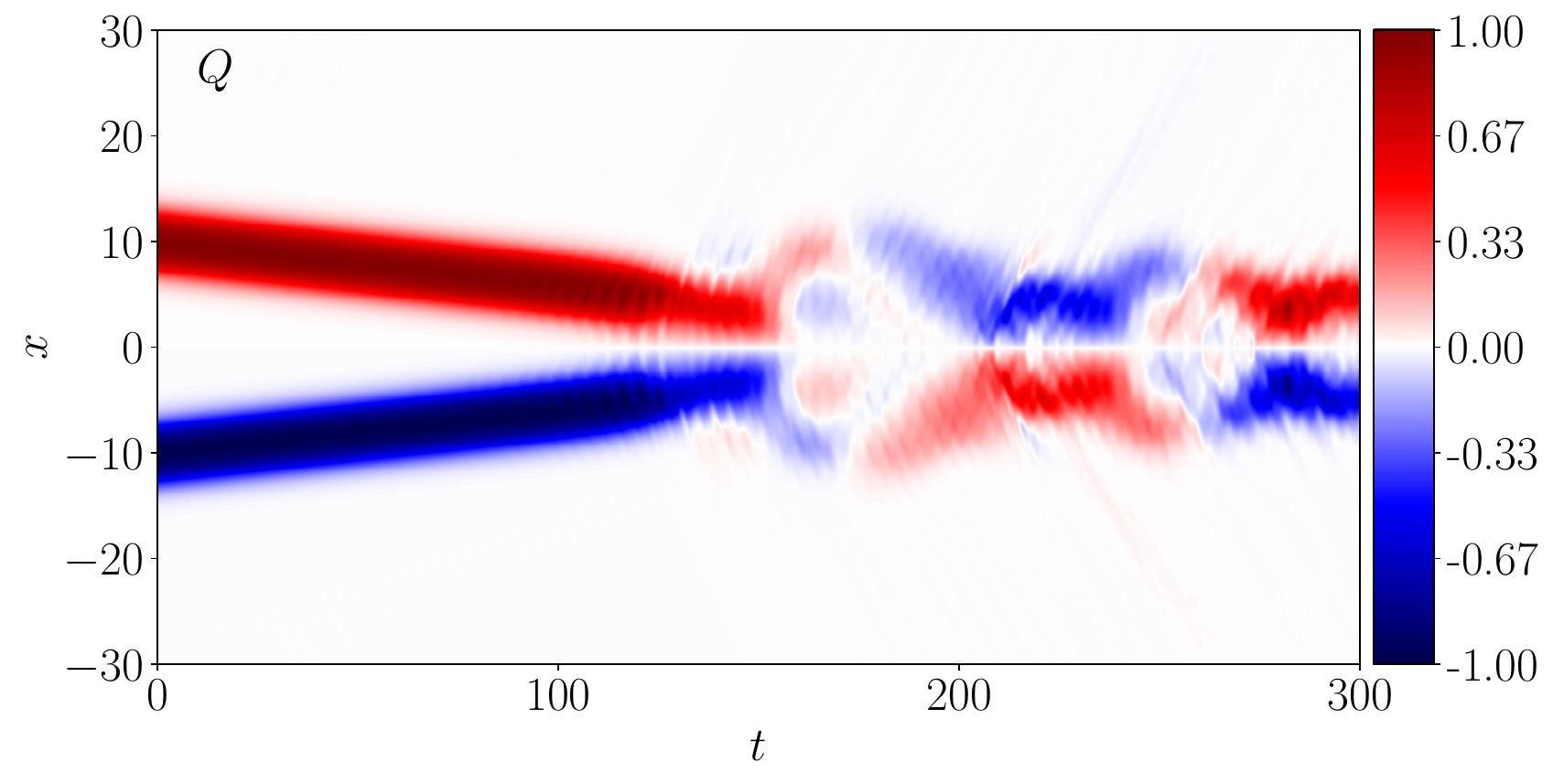}
 \includegraphics[width=0.50\textwidth]{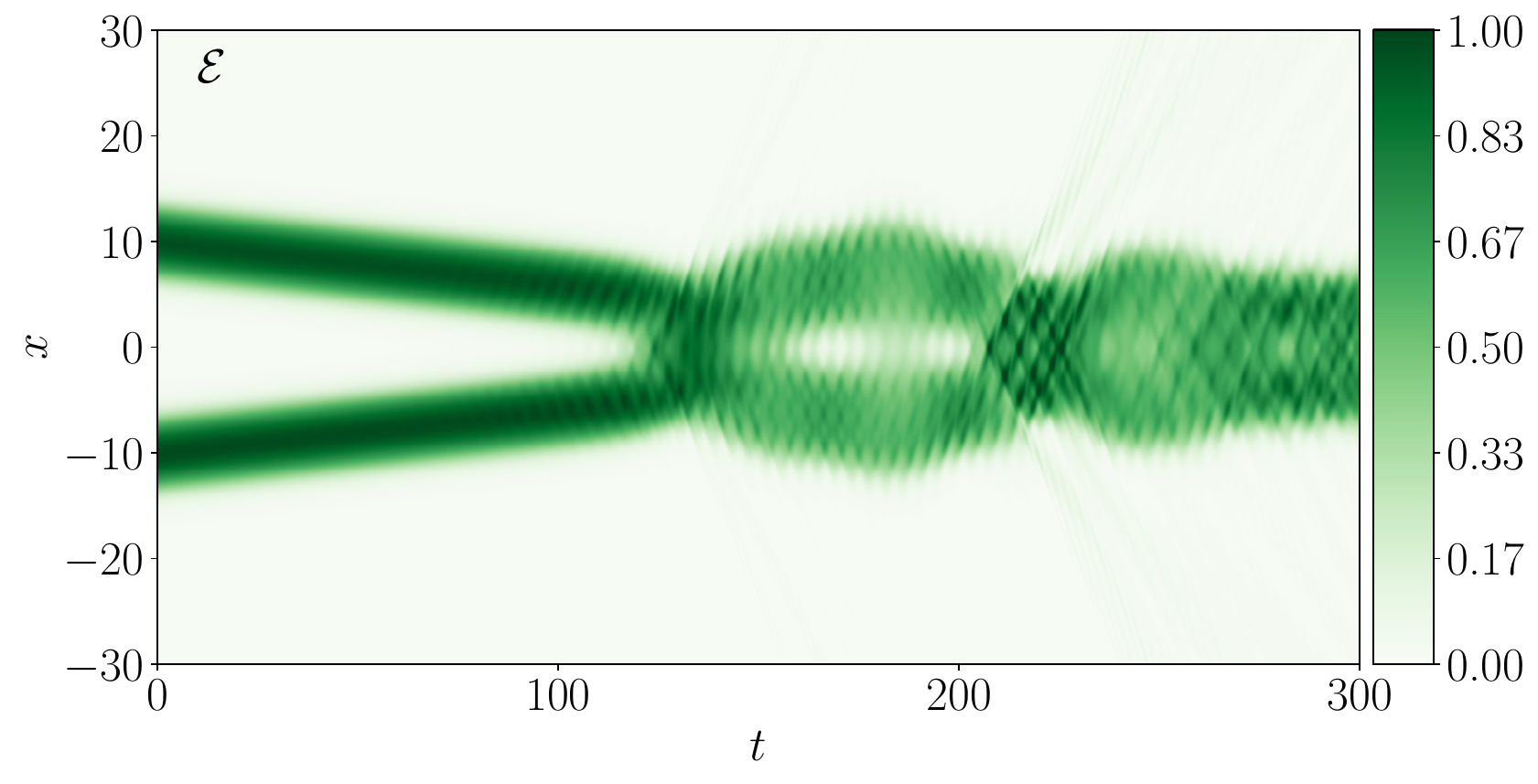}
  \includegraphics[width=0.50\textwidth]{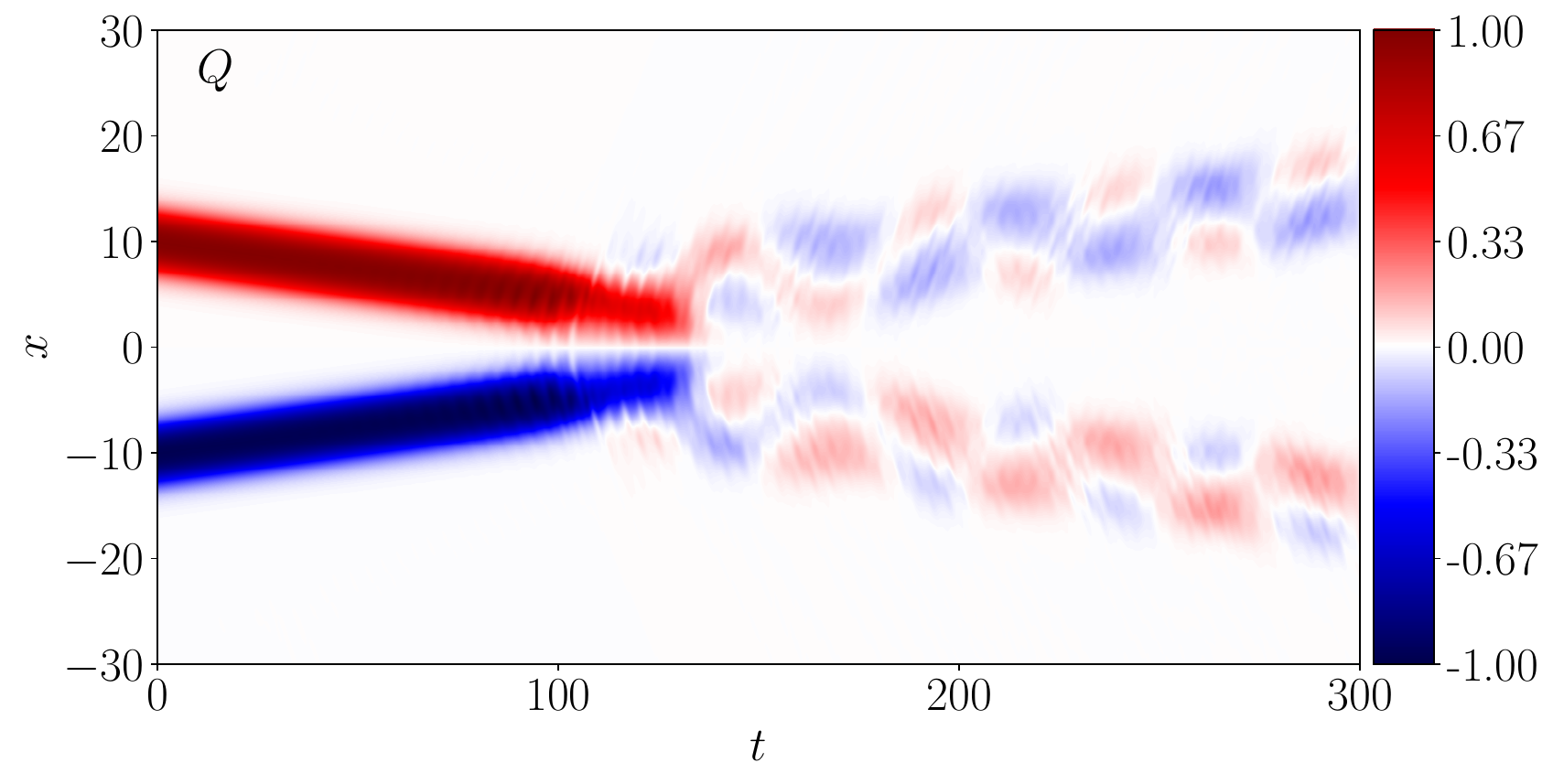}
 \includegraphics[width=0.50\textwidth]{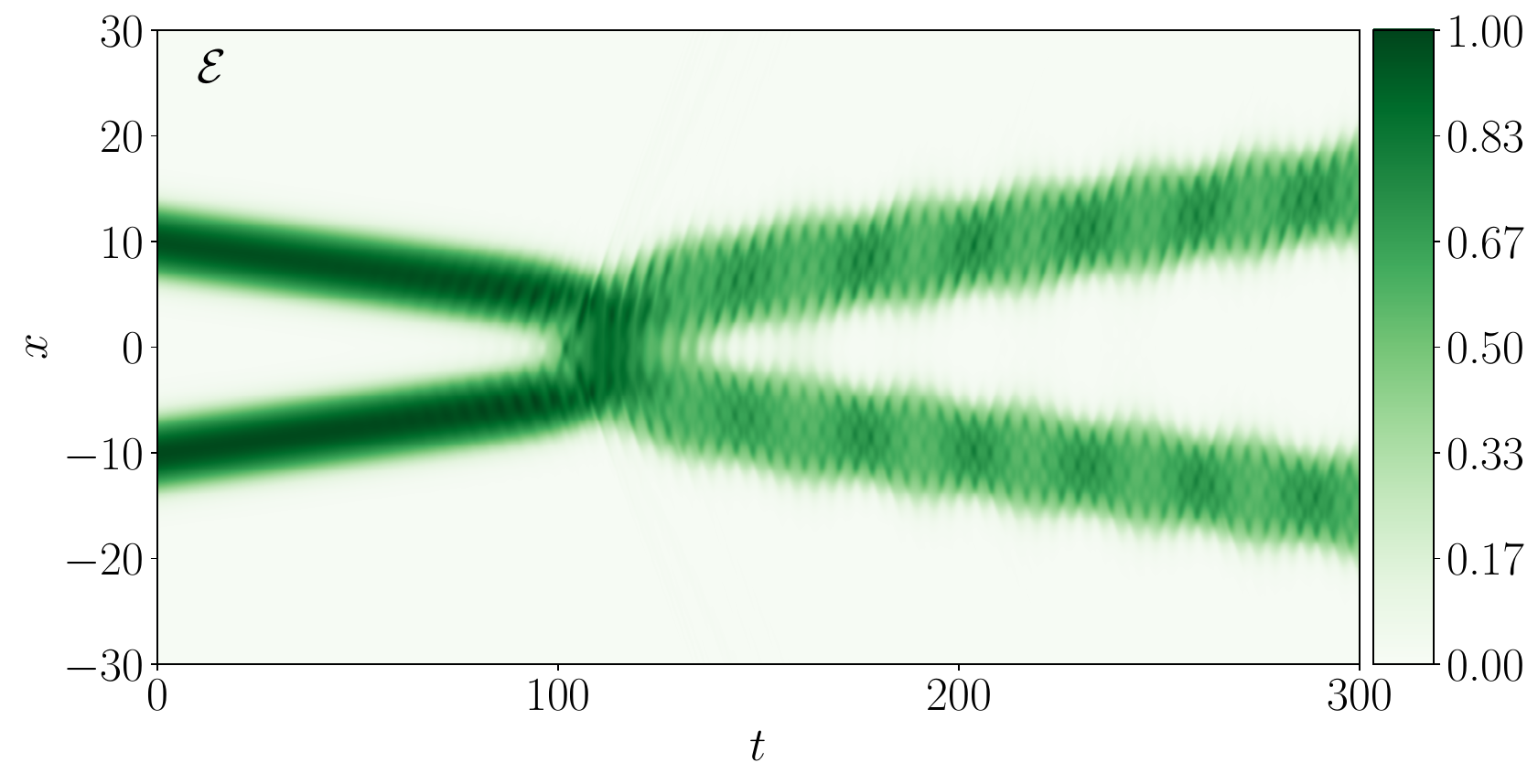}
 \includegraphics[width=0.50\textwidth]{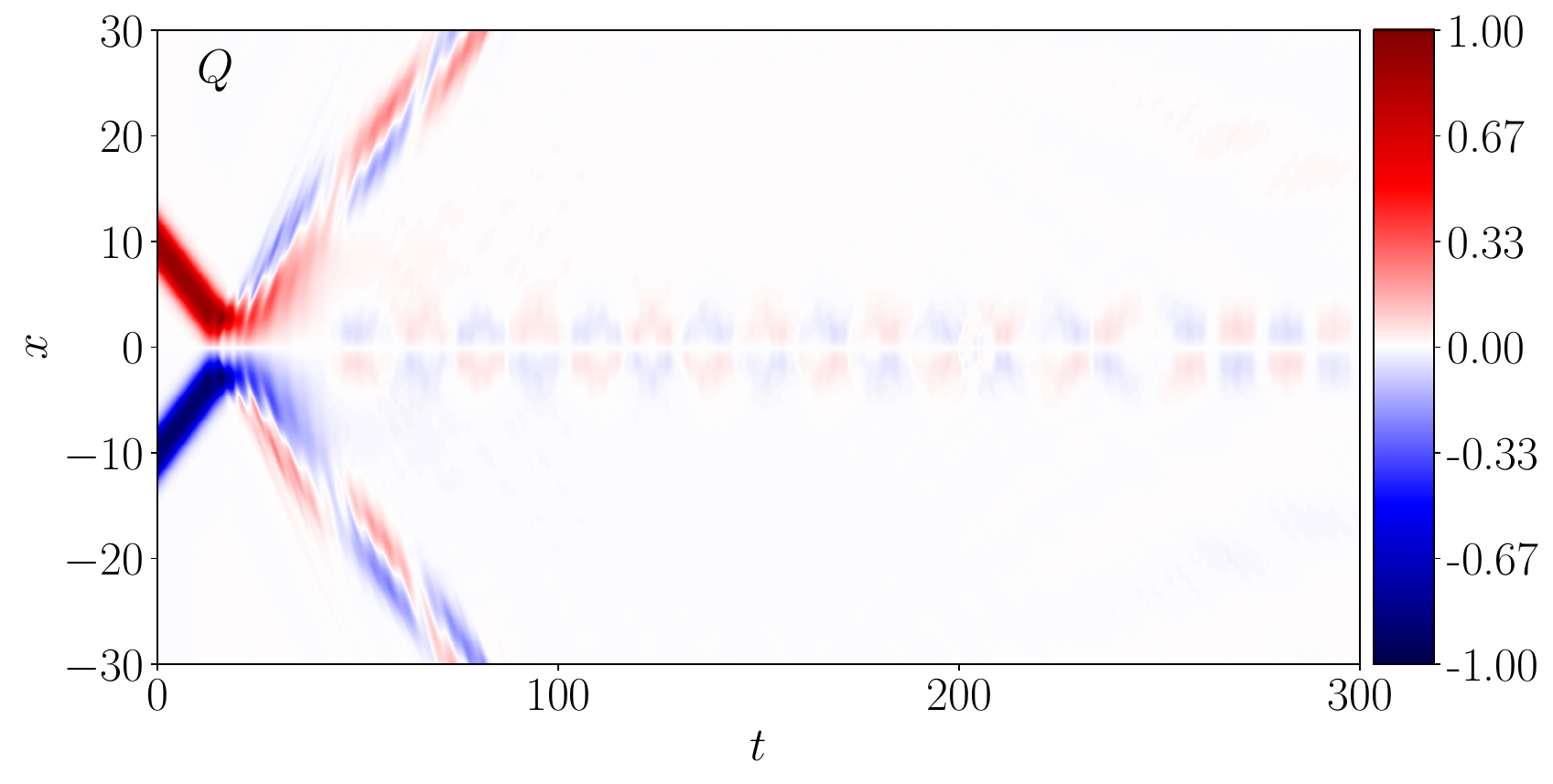}
 \includegraphics[width=0.50\textwidth]{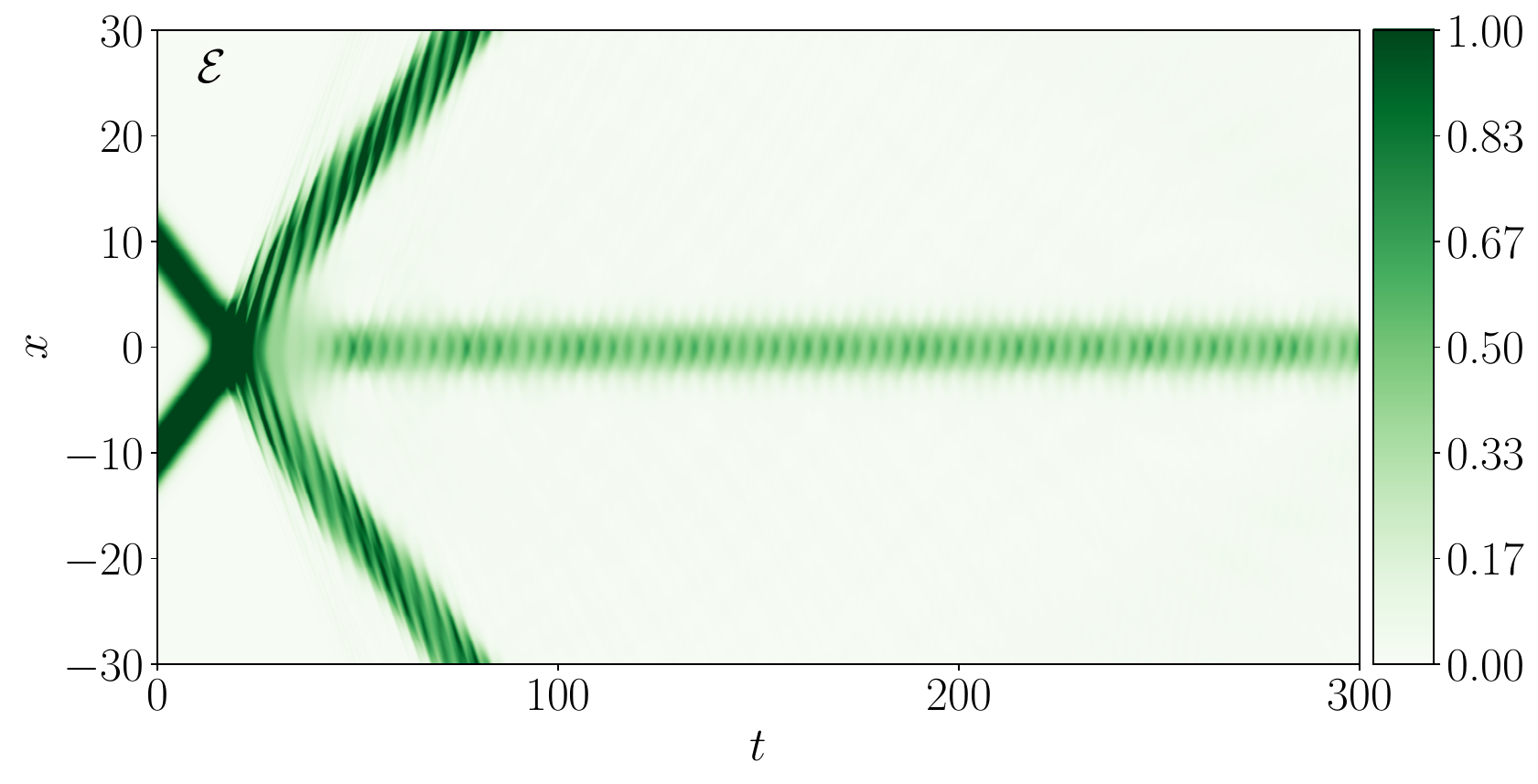}
 \caption{Charge density (right) and energy density of $QQ^*$ for $\beta=0.5$ and $\omega_0=\pm 0.708$. Here $v=0.04$ (upper panels), $v=0.05$ (central panels) and $v=0.5$ (lower panels). The initial phase is $\delta=0$.}
    \label{fig:no-bubble-1} 
\end{figure}

In Fig. \ref{fig:no-bubble} we present the energy density at the origin, i.e., at the point of the collision, for $\beta =0.5$ (i.e. $\omega_{min}=1/\sqrt{2}$) and $\omega_0=0.708$, and for initial velocity up to $0.7$.  This gives some information about the formation of any localized structure, as e.g., a polarized $Q$-ball state, at the origin (dark green) and about the ejection of perturbed $Q$-balls to infinity (white color at final $t$). Of course, one should be aware that both cases can occur in the same scattering process. 

There is a striking similarity to the corresponding plot for the kink-antikink collision in the $\phi^4$ model \cite{Sugiyama:1979mi, Campbell:1983xu}. There are very well visible self-similar structures for $v<0.06$, see Fig. \ref{fig:no-bubble} right panel, where we present a zoomed version. Here we have two scenarios. One-bounce or even two-bounce windows, where the incoming $Q$-balls collide once or twice, and then they reappear in the final state. Note that the ejected solitons are typically perturbed polarized $Q$-balls. E.g., each of them can be a charge-swapping state (each with non-zero total charge), see the top panels of Fig. \ref{fig:no-bubble-1}.

In the second scenario, the $Q$-balls form a sort of bion-type solution. Here, they never show up as independent polarized $Q$-balls but perform a huge number of bounces and eventually, passing through an oscillon state, decay to the vacuum, see  the central panels of Fig. \ref{fig:no-bubble-1}. These collisions are also very similar to scatterings of oscillons, where almost identical structures are formed \cite{Blaschke:2024uec}. 

For $v\in [0.06,0.25]$ there is a large one-bounce scattering region with two polarized $Q$-balls in the final state. Although one may think this is a simple regime, that is not really the case. We find various polarized charge-swapping solutions with different total charges. As the initial velocity increases, $v\in [0.25,0.65]$, the third remnant of the collision emerges. It is another polarized $Q$-ball, here a charge-swapping state, created at the origin. For $v>0.65$ there are four polarized $Q$-balls produced after the first collision. Finally, for $v\approx 0.9$ the $Q$-balls pass through each other.

%%%%%%%%%%%%%%%%%%%%%%%%%%%%%%
\subsection{The bubble in $Q$-ball dynamics}
%%%%%%%%%%%%%%%%%%%%%%%%%%%%%%
\begin{figure}%[h!]
 \includegraphics[width=0.55\textwidth]{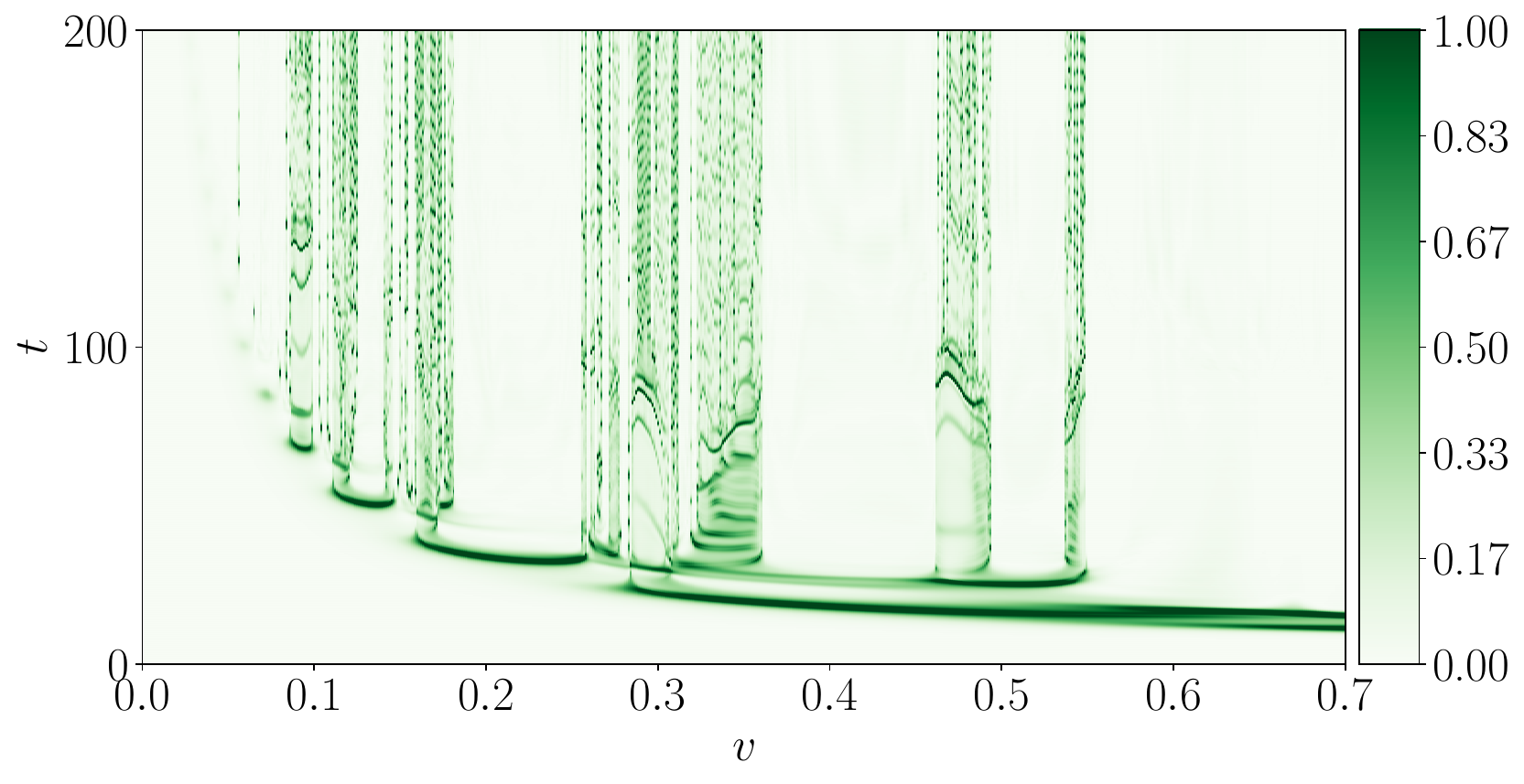}
 \includegraphics[width=0.55\textwidth]{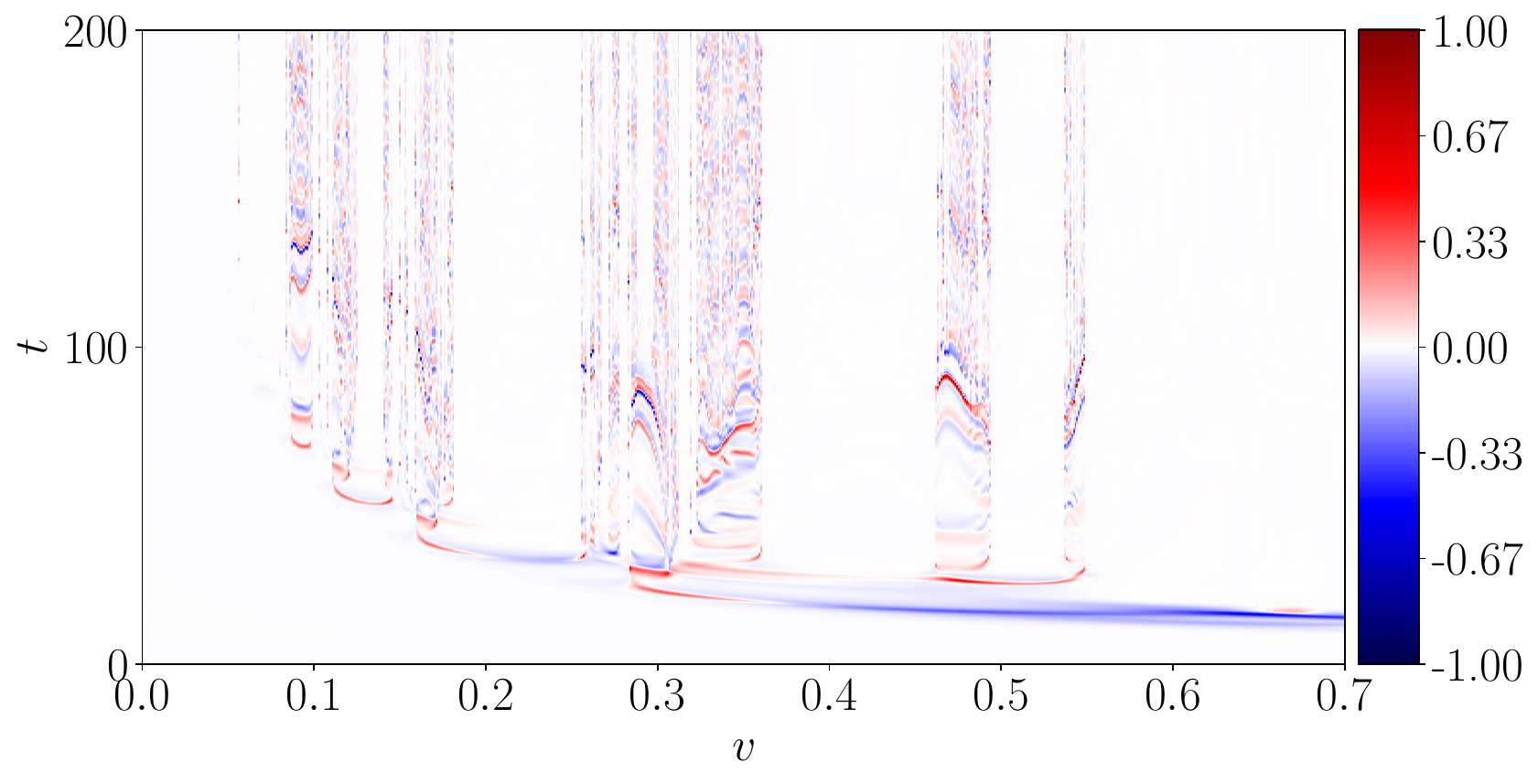}
 \caption{Evolution of the energy density $\mathcal{E}(x=0,t)$ (left) and charge density $j^0(x=0.05,t)$ in time for different values of the initial velocity of the $Q$-ball. Here, $\beta=0.26$ and incoming $Q$-balls are in the thin-wall regime, $\omega_0=\pm 0.2$. The initial phase is $\delta=0$.}
    \label{fig:QQ*-v} 

    \vspace*{0.5cm}
    
 \includegraphics[width=0.55\textwidth]{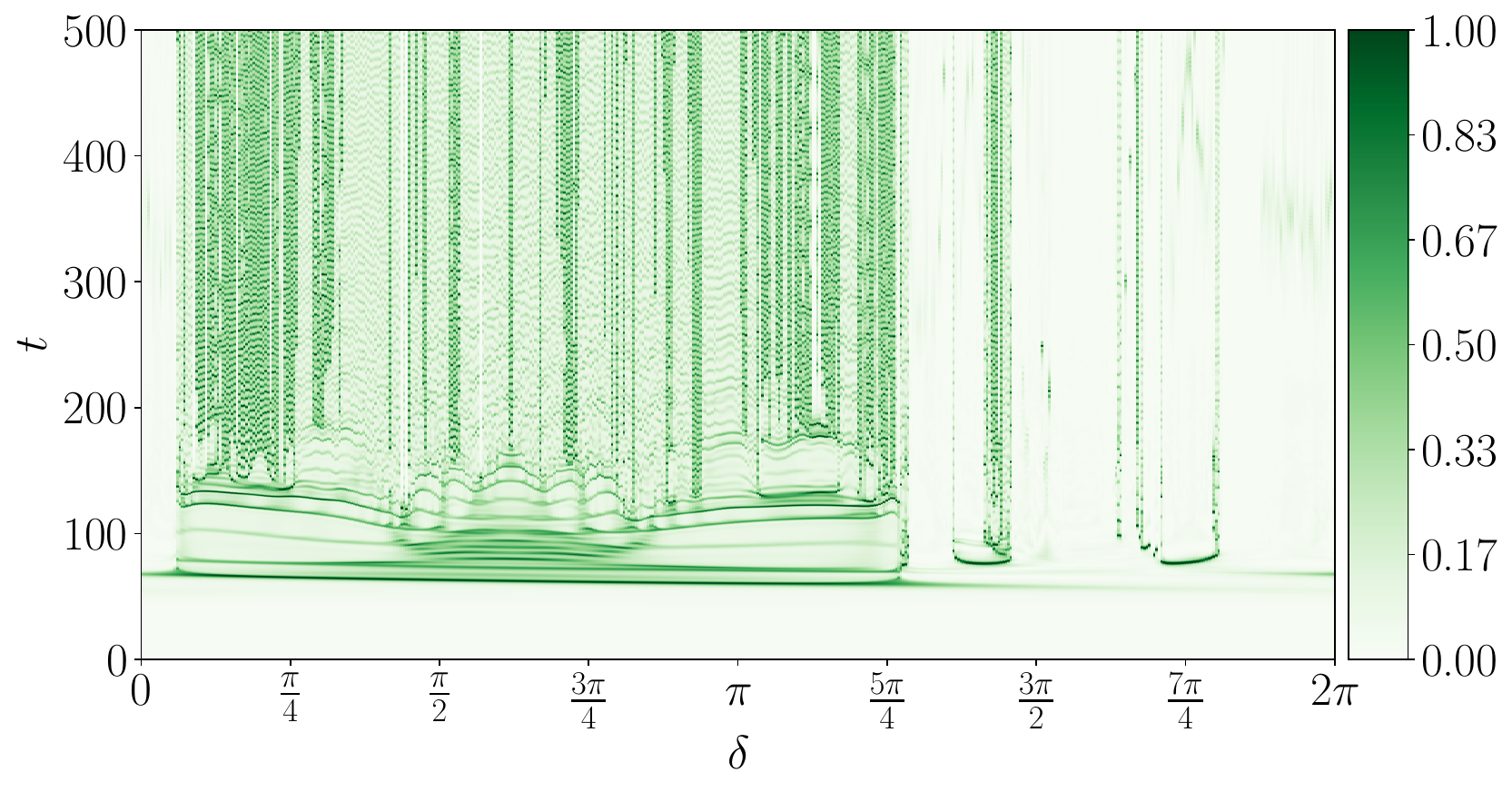}
 \includegraphics[width=0.55\textwidth]{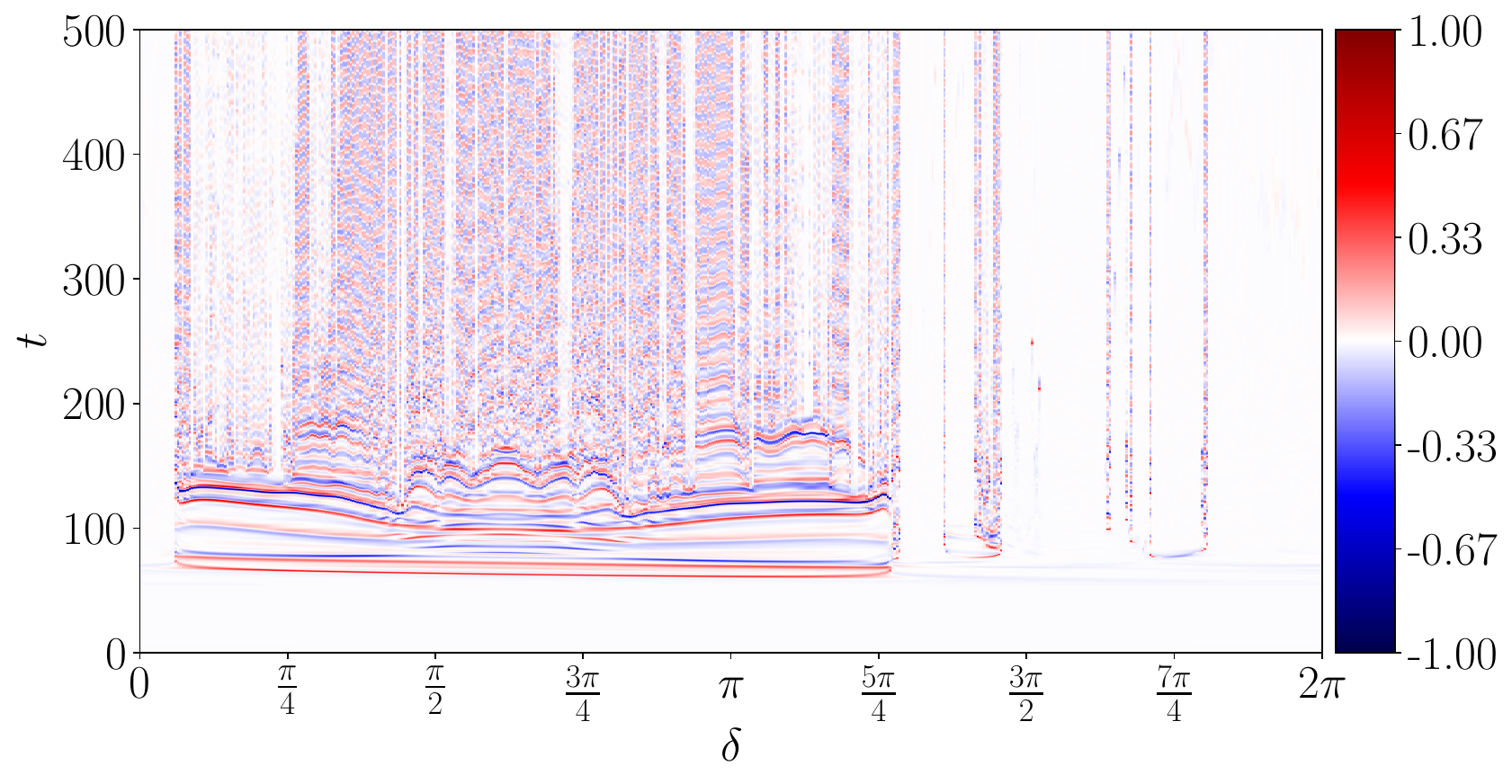}
    \caption{Evolution of the energy density $\mathcal{E}(x=0,t)$ (left) and charge density $j^0(x=0.05,t)$ in time for different values of the initial phase of the $Q$-ball. Here, $\beta=0.26$ and incoming $Q$-balls are in the thin-wall regime, $\omega_0=\pm 0.2$. The initial velocity is $v=0.1$.}
    \label{fig:QQ*-delta}  
\end{figure}

Now we consider the case where the potential has a false vacuum, and so has a bubble solution. Specifically, we assume $\beta=0.26$.  In Fig. \ref{fig:QQ*-v} we show the $QQ^*$ dynamics for the thin-wall regime, $\omega_0=0.2$. We plot the time evolution of the energy density $\mathcal{E}$ at the center of mass and the time evolution of the charge density $j^0$ at $x=0.05$ (at $x=0$ the charge density is always zero due to symmetry). In both cases, we scan over the initial velocity assuming that $\delta=0$. 

It is clearly visible that the $QQ^*$ collision reveals a chaotic, self-similar pattern, which, however, differs significantly from the previous case, Fig. \ref{fig:no-bubble}. The difference is due to the appearance of the bubble as an intermediate state, which definitely contributes to a further complexity of the scattering.  Once again, there are single collision scenarios separated, in a chaotic way, by regions where at the origin a long-lived charged state is formed.  A very similar pattern is shown in Fig. \ref{fig:QQ*-delta}, where we fix the initial velocity, $v=0.1$, but change the relative phase of the incoming $Q$-balls.

\begin{figure}%[h!]
 \includegraphics[width=1.00\textwidth]{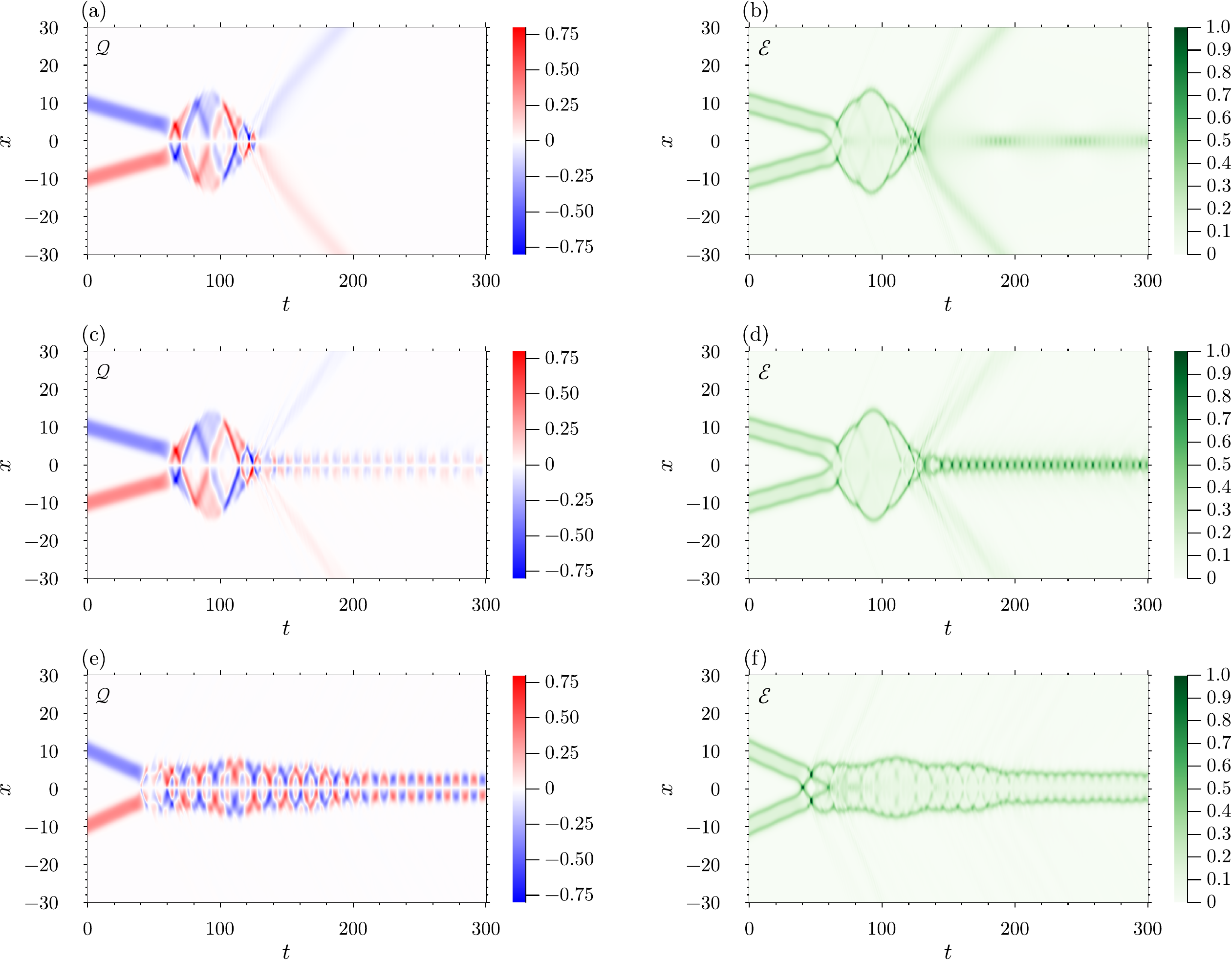}
 % \includegraphics[width=1.00\textwidth]{frame0097.png}
    % \centering2
    %\includegraphics[width=1.00\textwidth]{QE_delta_0.1pi-crop.pdf}
    % \includegraphics[width=1.05\textwidth]{QE_delta_0.25pi_v0.16.pdf}
    \caption{Examples of $QQ^*$ collisions where the bubble solution hosting the Goldstone waves is produced. Here, $\beta=0.26$, $\omega_0=0.2$. The initial conditions are $v_{in}=0.1$ $\delta=1.07\pi$ (a,b), $\delta=1.1\pi$ (c,d) and $v_{in}=0.16$ $\delta=1.25\pi$ (e,f). }
    \label{fig:QQ*-bubble}  
\end{figure}

Now we look closer at particular examples of the $QQ^*$ collisions and clarify the role of the bubble. We found that the bubble very often appears as an intermediate state, see Fig. \ref{fig:QQ*-bubble}. In fact, it is the first state that appears after the $Q$-balls meet for the first time and temporarily annihilates. However, because each $Q$-ball has a nontrivial charge density, the bubble is always produced in an excited state, that is, with the Goldstone modes. We can see that the charge of each $Q$-ball delocalizes in the bubble in the form of ``free" Goldstone waves. As the waves scatter from the bubble boundaries, they blow the bubble up. Then, depending on the particular conditions, the bubble can collapse or temporarily stabilize.  

If the bubble collapses, the charge can be re-confined into localized objects. This is presented in Fig. \ref{fig:QQ*-bubble} in the top and central panels, where, after the collapse of the bubble, a $Q$-ball and anti-$Q$-ball reappear, ejected with a relatively high velocity, approximately four times larger than the initial velocity. A significant fraction of the charge is annihilated during the creation and collapse of the bubble. Hence, the final $Q$-balls carry a smaller value of the $U(1)$ charge, that is, they rotate with higher internal frequency. For the upper panel, the final $Q$-balls have $\omega_0 \approx 0.94$. Interestingly, there is another remnant of the collision, an oscillon, staying at the center of mass. Typically, the oscillon is created in an excited state. In Fig. \ref{fig:QQ*-bubble} upper panel, the excitation concerns only the real component of the field and we find a modulated oscillon which has zero $U(1)$ charge. In Fig. \ref{fig:QQ*-bubble} central panel, the oscillon is excited in the imaginary component of the field, and therefore is a member of the polarized $Q$-ball family, which leads to the well-known Copeland-Saffin-Zhou (CSZ) charge-swapping solution. Of course, these oscillons are not ultimately stable states. E.g., after a long time, the charge-swapping solution decays to an oscillon with trivial charge density \cite{Copeland:2014qra}. 

\begin{figure}%[h!]
    \includegraphics[width=1.00\textwidth]{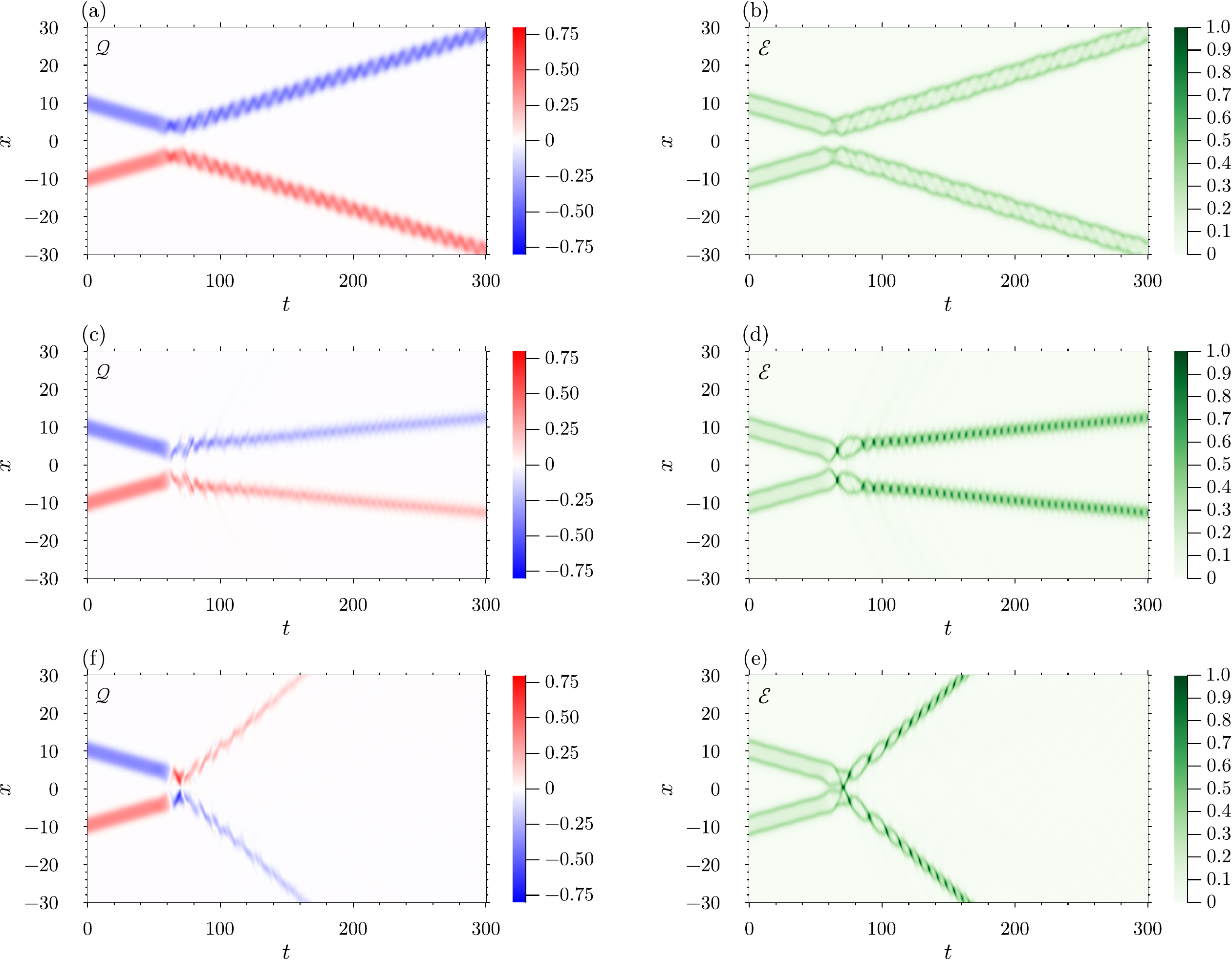}
     \caption{No annihilation scenarios in the $QQ^*$ collisions. Here $\beta=0.26$ and $v_{in}=0.1$, $\omega_0=0.2$.  Upper panel: $\delta=1.21 \pi$ and two charged oscillons in the final state. Central panel: $\delta=1.69 \pi$ and two excited $Q$-balls. Lower panel: transition scenario with $\delta=1.13 \pi$.}
    \label{fig:QQ*-noani}  
\end{figure}
\begin{figure}%[h!]
      \includegraphics[width=1.00\textwidth]{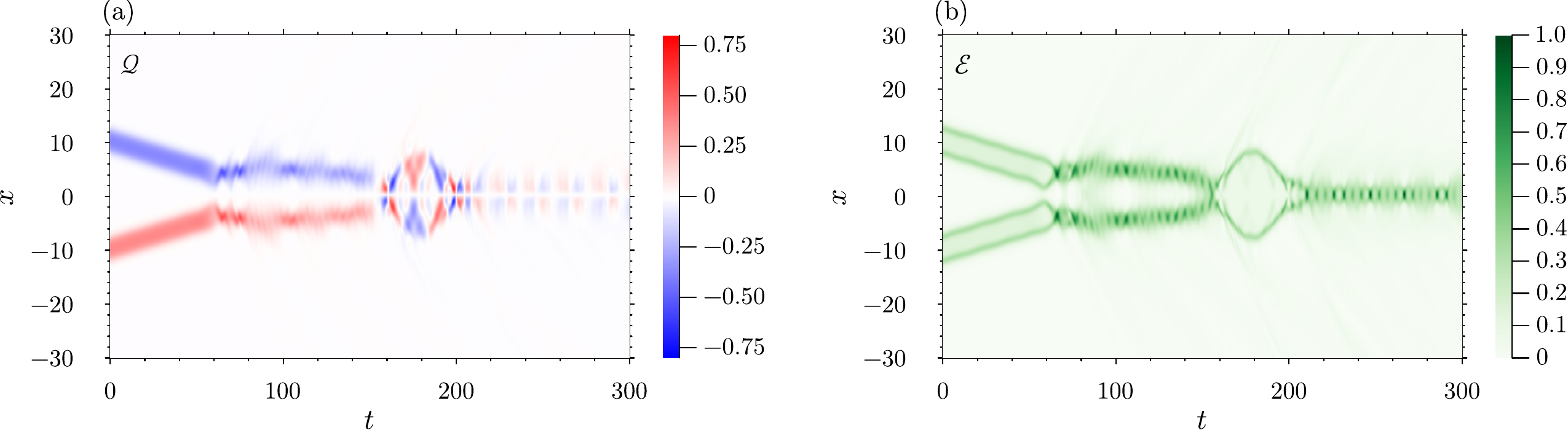}
     \caption{Example of the creation of the bubble in second bounce. Here $\beta=0.26$, $v_{in}=0.1$, $\omega_0=0.2$ and $\delta=1.37 \pi$.}
    \label{fig:QQ*-3}  
\end{figure}

In some cases, the bubble has a smaller size and evolves in a more stable way. After the initial intermediate stage, it stabilizes to a fixed size. This is a fairly long-lived solution with a charge-swapping property, which is of a {\it different origin from the CSZ charge-swapping}. It is triggered by the Goldstone wave trapped in the bubble, not by a perturbed oscillon, see Fig. \ref{fig:QQ*-bubble} lower panels. Specifically, in this case, the Goldstone waves excite the first trapped mode.  

In the collision, the $Q$-ball and anti-$Q$-ball do not have to necessarily form the bubble. They can keep their identity for the whole evolution. This leads to a more or less elastic scattering or to a scenario in which the $Q$-balls pass through each other. The first possibility is presented in Fig. \ref{fig:QQ*-noani}, the upper and central panels. Specifically, in the upper panels ($v_{in}=0.1$, $\omega_0=0.2$ and $\delta=1.21 \pi$) the colliding $Q$-balls repel each other. This results in a back-scattered pair of $Q$-balls. Note that the collision excites the internal (vibrational) modes of the $Q$-balls. 
In the central panels ($v_{in}=0.1$, $\omega_0=0.2$ and $\delta=1.69 \pi$) the repulsion is weaker, which leads to a partial annihilation of the charge and to stronger excited $Q$-balls. Then, they very quickly relax to charged oscillons. Again, we expect that the oscillons will eventually decay to stable $Q$-balls. In Fig. \ref{fig:QQ*-noani}, the lowest panel ($v_{in}=0.1$, $\omega_0=0.2$ and $\delta=1.13 \pi$), the charge densities pass through each other. Here, after the collision, the initial $Q$-balls relax to perturbed charged oscillons. Their velocity is much larger than the initial velocity of the $Q$-balls. We note that the perturbed oscillon may also be in the CSZ charge-swapping state. This occurs for $\delta=1.15 \pi$.

Although bubbles most commonly form in the initial collision of Q-balls, they may also form in the collision of the initially-created charged oscillons, see Fig. \ref{fig:QQ*-3}. Here, $v_{in}=0.1$, $\omega_0=0.2$ and $\delta=1.37 \pi$. This is not a surprising scenario since $Q$-balls and oscillons (together with their perturbed states) are members of the family of polarized states and therefore can be smoothly transformed into the one another, if a suitable perturbation is added.

It is clear that $QQ^*$ collisions in the thin-wall regime in the $\phi^6$ model with a well-pronounced false vacuum reveal a highly complex structure with a plethora of scenarios. The actual evolution is extremely sensitive to the initial condition, e.g., the initial velocity and phase. Even a tiny change in the initial data can lead to a drastic change in the actual dynamics and in the final state formation. This gives further support for the existence of chaos in such collisions.

%%%%%%%%%%%%%%%%%%%%%%%%%%%%%%
\subsection{The sphaleron in $Q$-ball dynamics}
%%%%%%%%%%%%%%%%%%%%%%%%%%%%%%
\begin{figure}%[h!]
\includegraphics[width=0.5\textwidth]{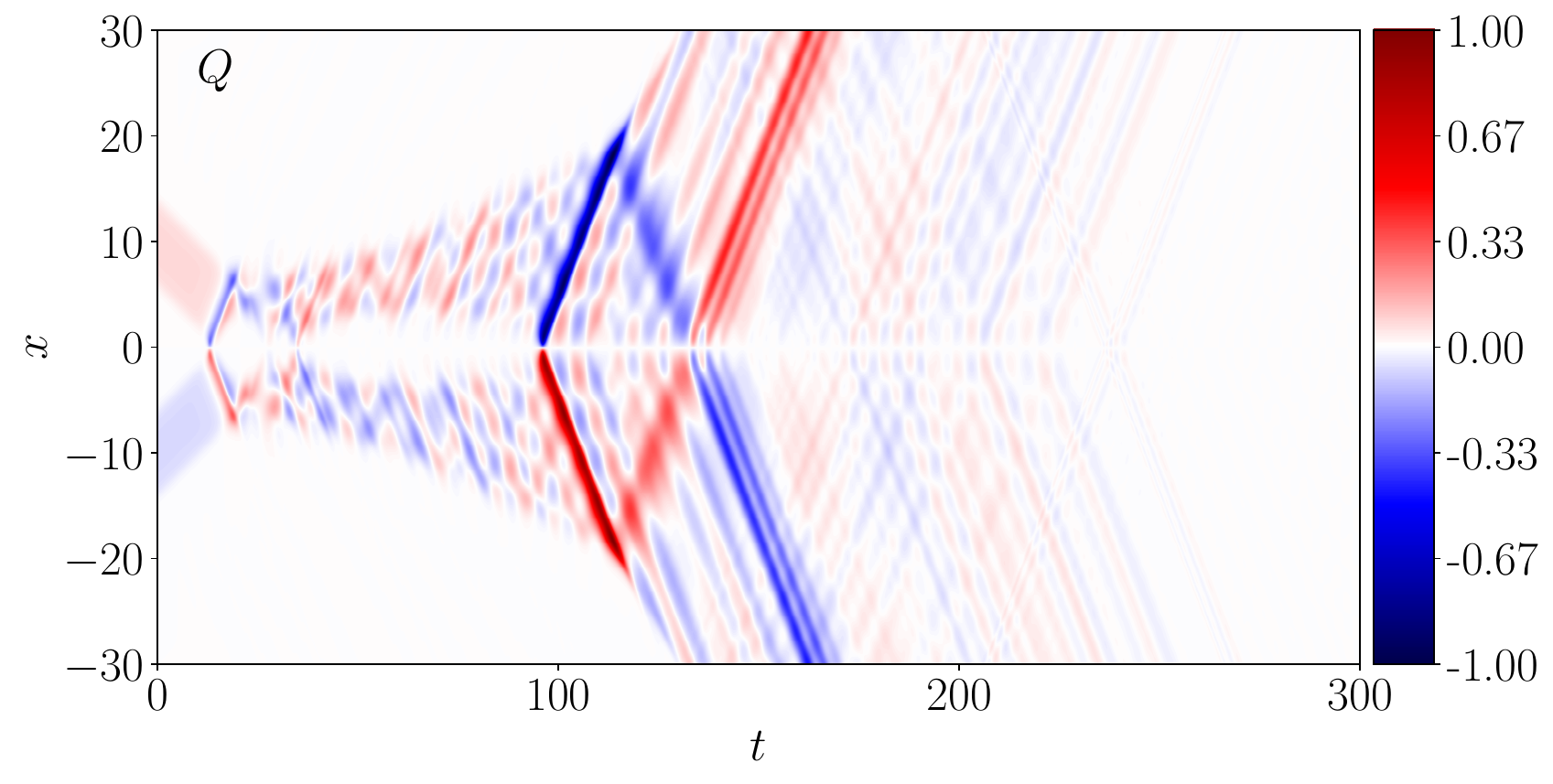}
    \includegraphics[width=0.5\textwidth]{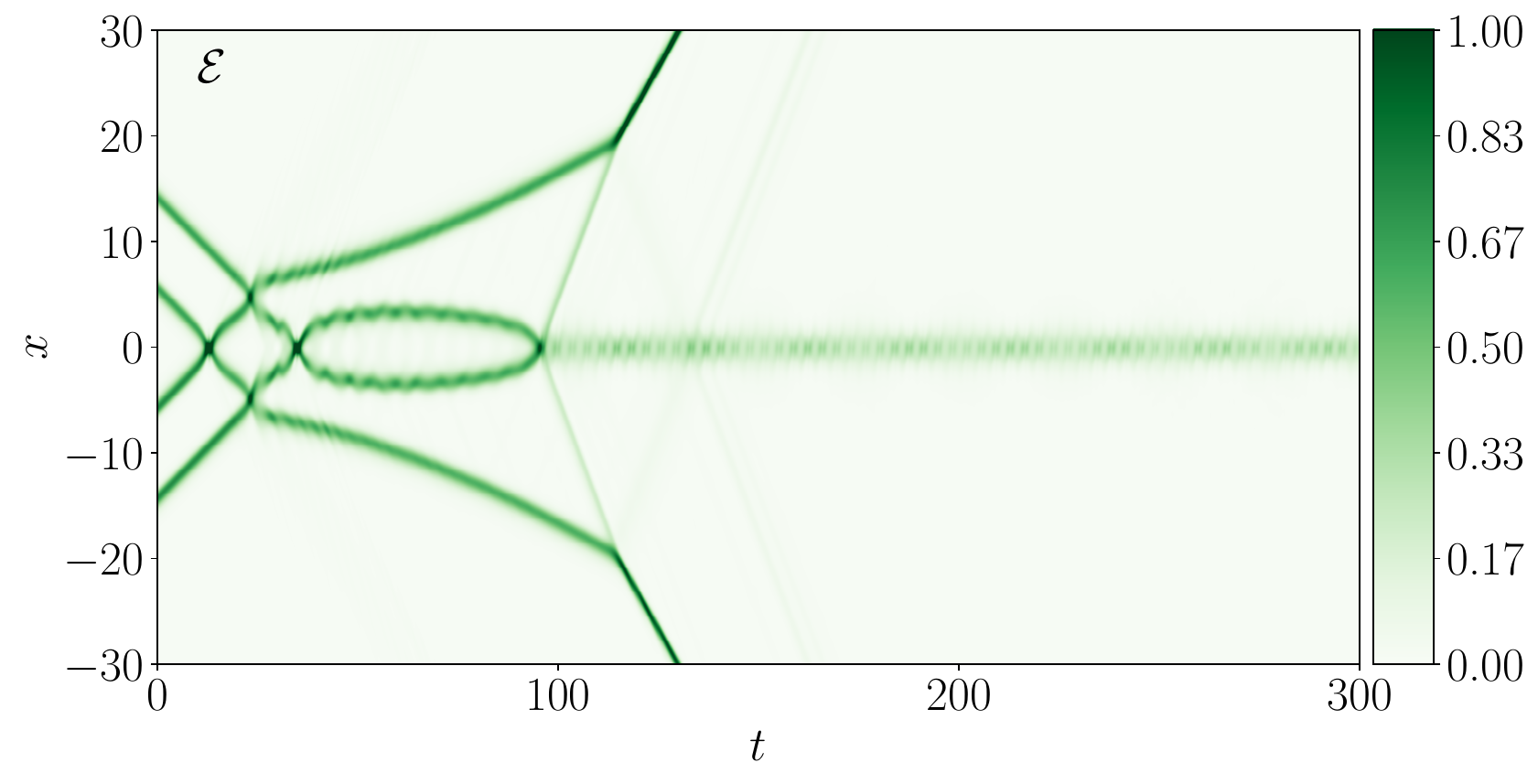}
    \caption{Example of $QQ^*$-ball collision with the sphaleron as an intermediate state. Here $\beta=0.2501$ and $\omega=0.02$ and the initial velocity is $v_{in}=0.3815$.}
    \label{fig:S}   

\vspace*{0.2cm}
    
    \includegraphics[width=0.5\textwidth]{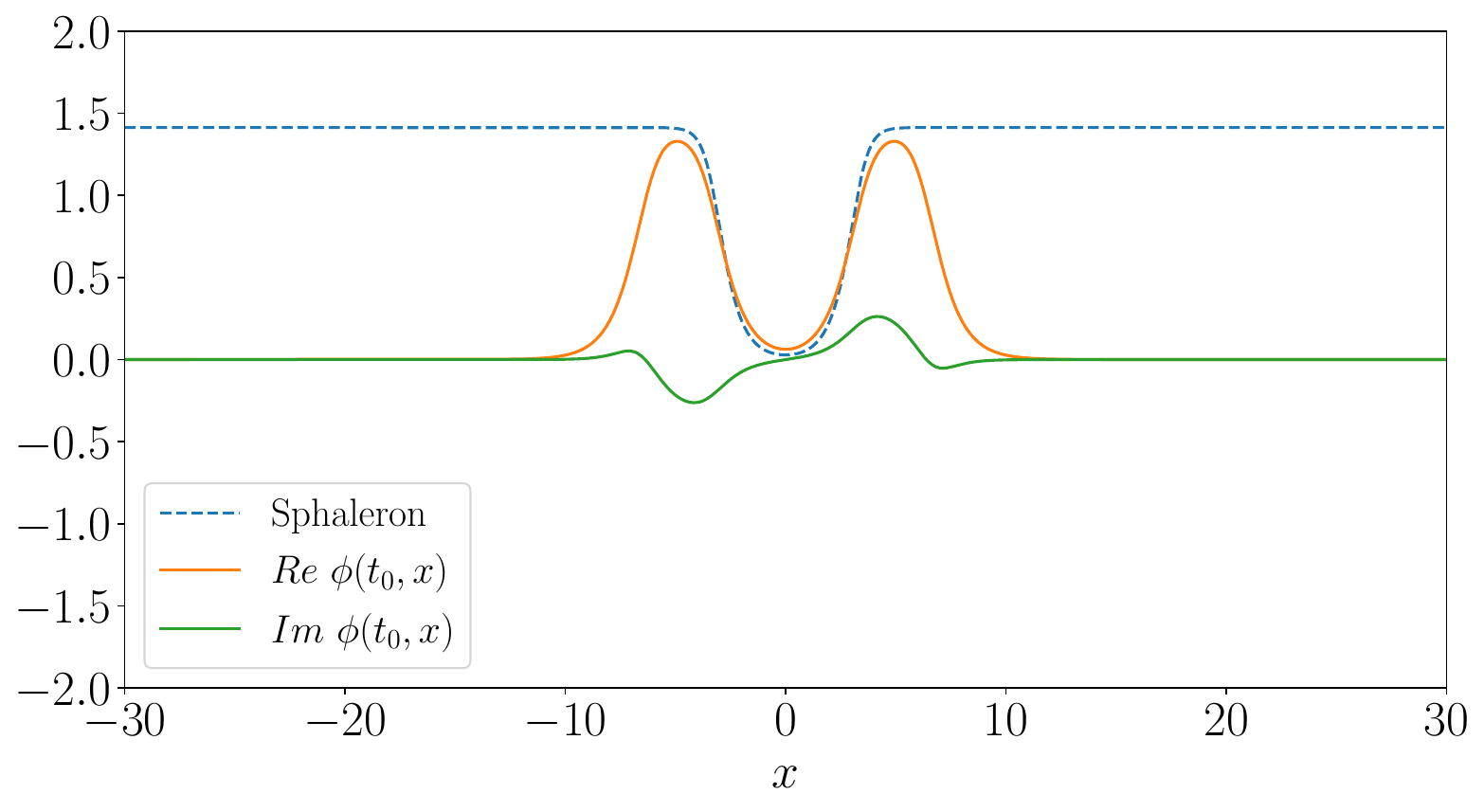}
    \includegraphics[width=0.5\textwidth]{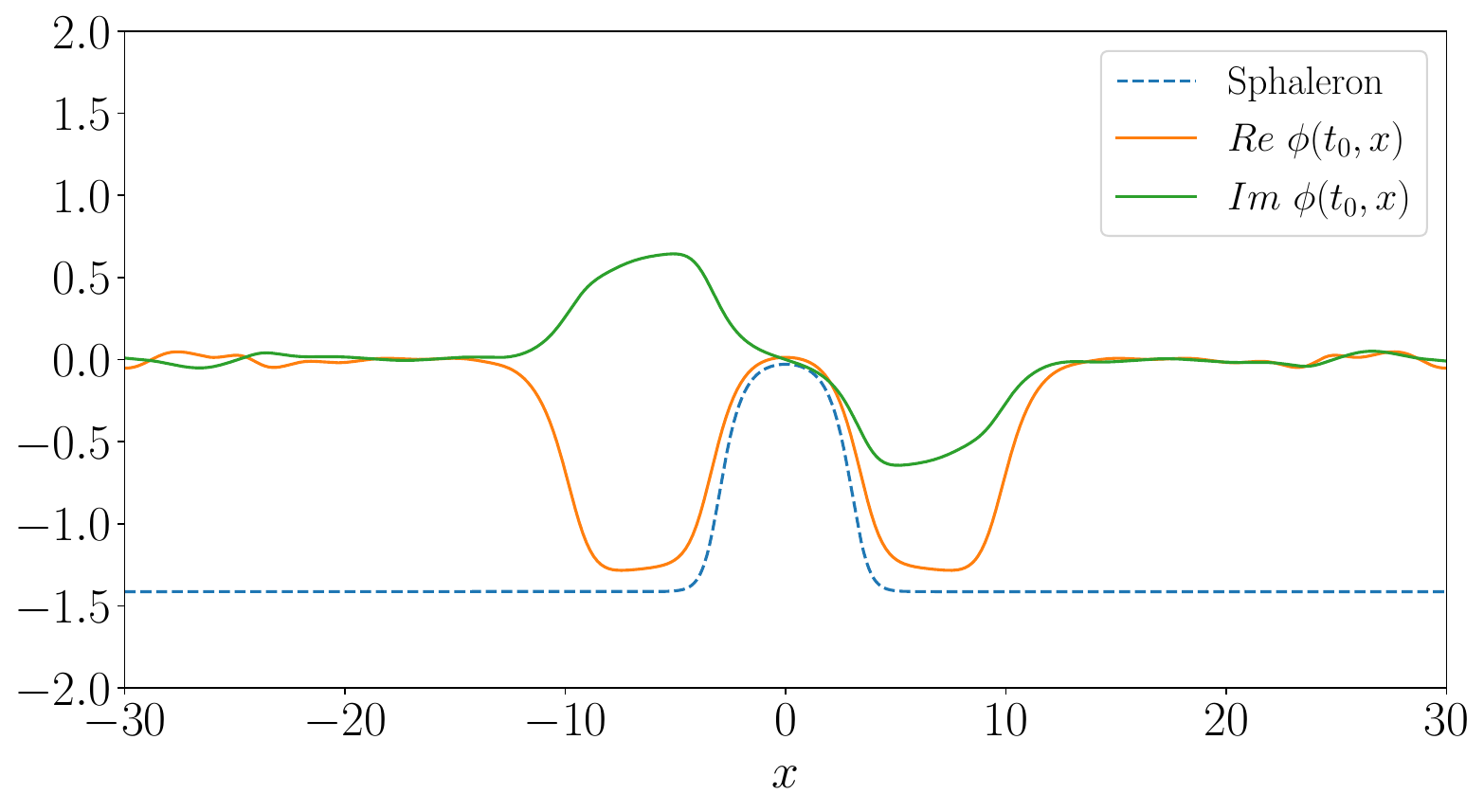}
     \caption{The real (orange) and imaginary (green) component of the field at $t=20$ and $t=60$ with the sphaleron dynamically formed at the origin. The pure sphaleron solution is plotted in blue.}
     \label{fig:S-profile}   
\end{figure}
As we explained, the sphaleron exists as a stationary point for which the (real component of the) field asymptotes to the false vacuum value. Therefore, it can emerge in the $QQ^*$ collisions as an excitation inside of the bubble. Obviously, once $\beta$ is closer to the critical value $1/4$, the sphaleron is more clearly visible. Once created, it can decay into the false vacuum oscillon, the bubble, or into the true vacuum. 

In Figs \ref{fig:S} and \ref{fig:S-profile} we present an example of the creation of the sphaleron in the $QQ^*$ collision. Here, $\beta=0.2501$ (i.e. $\omega_{min}=0.01996)$ and $\omega_0=0.02$. The initial velocity of the  $Q$-balls is $0.3815$.  The sphaleron is formed twice, in the first and second collisions. The second sphaleron lives for a relatively long time $\Delta t \sim 50$ which can be an effect of the excitation of its mode \cite{Navarro-Obregon:2024ieb, Navarro-Obregon:2025xmw}, see the wiggles in Fig. \ref{fig:S}, right panel. Then, it decays into the false vacuum oscillon inside the bubble full of Goldstone waves. 

%%%%%%%%%%%%%%%%%%%%%%%%%%%%%%
\section{Collision of the $Q$-ball with the bubble}
%%%%%%%%%%%%%%%%%%%%%%%%%%%%%%

As we have seen, the bubble easily appears in $QQ^*$ collisions, whenever the potential possesses a broken false vacuum. Such a bubble is created with a bunch of the excited Goldstone modes and, in a more complicated multi-$Q$-ball environment,  may further interact with other $Q$-balls. Here we briefly study the simplest case, which is the interaction of a $Q$-ball with an unexcited bubble.

\begin{figure}%[h!]
 \centering
\includegraphics[width=0.95\textwidth]{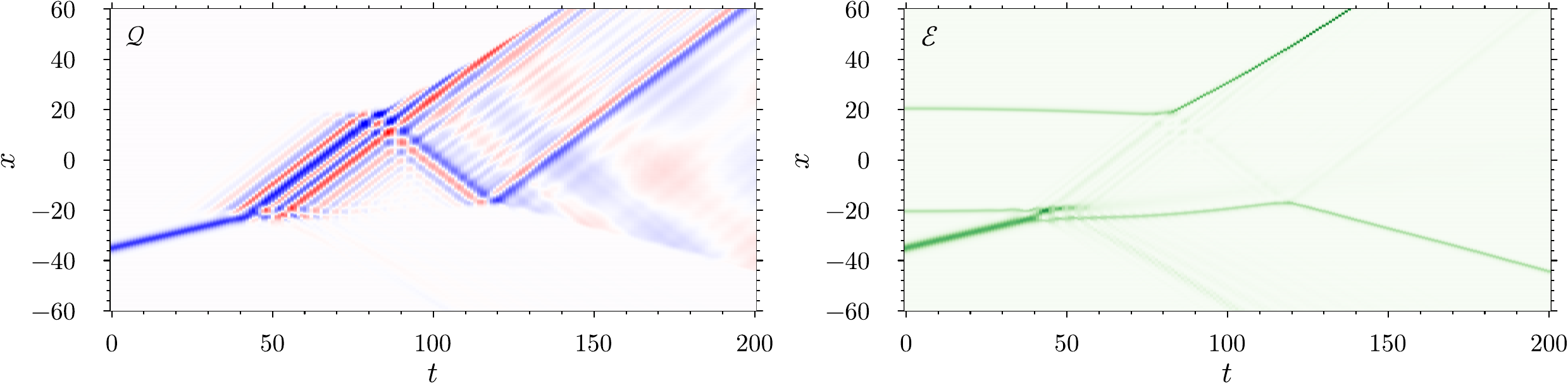}
    \includegraphics[width=0.95\textwidth]{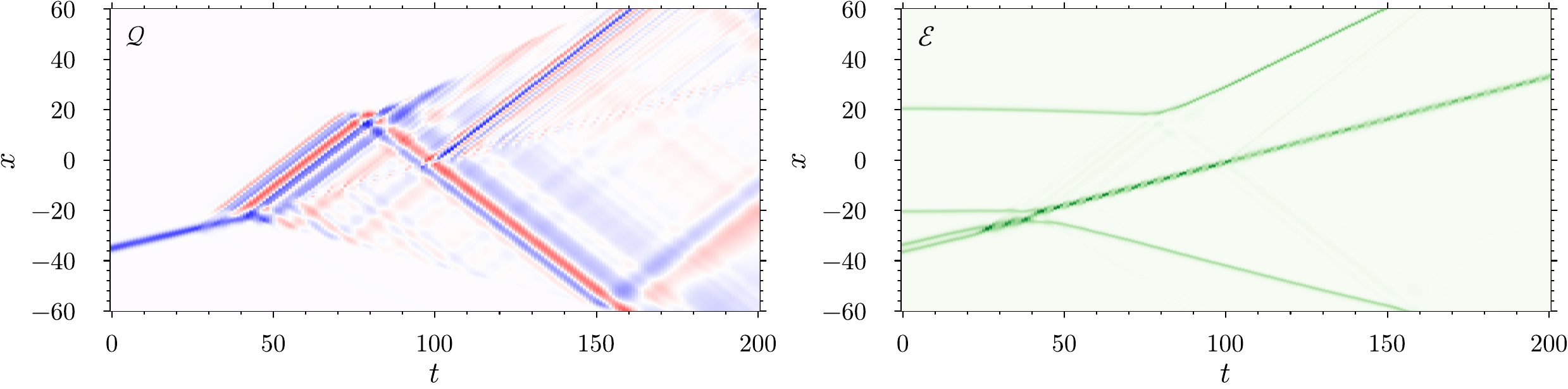}
    \caption{Examples of $Q$-ball bubble collisions. Here $\beta=0.2501$ and $\omega=0.75$ (top) and $\omega=0.2$ (bottom). Initial velocity is $v_{in}=0.3$.}
    \label{fig:QB-1}       
 \centering

\vspace*{0.2cm}

\includegraphics[width=0.95\textwidth]{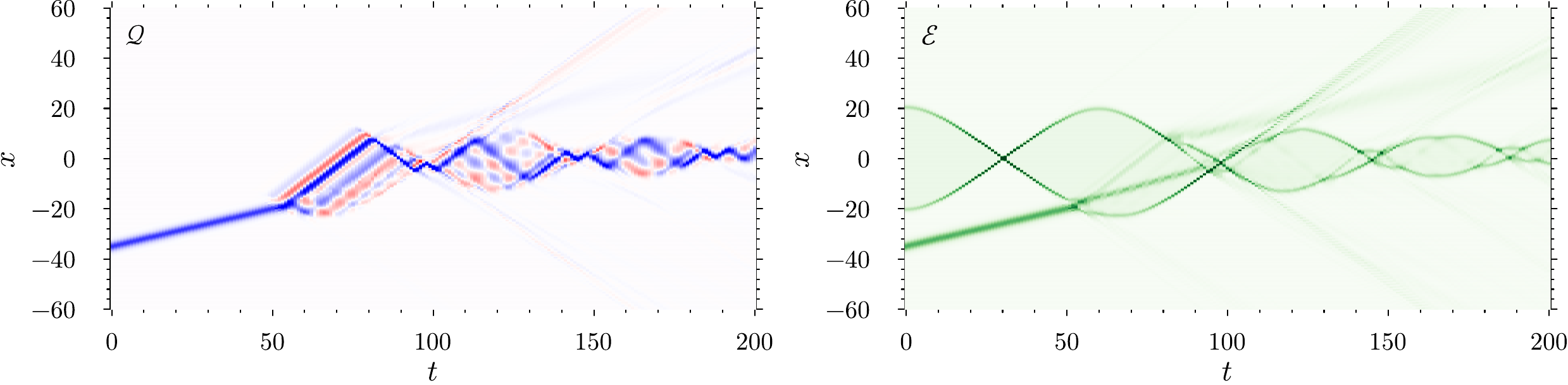}
    \includegraphics[width=0.95\textwidth]{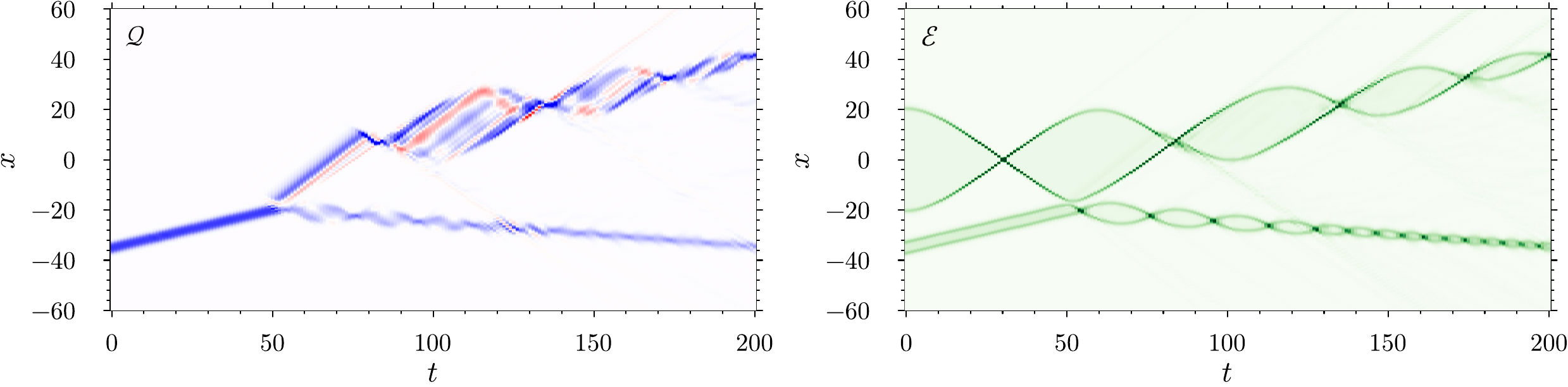}
    \caption{Examples of $Q$-ball bubble collisions. Here $\beta=0.26$ and $\omega=0.75$ (top) and $\omega=0.2$ (bottom). Initial velocity is $v_{in}=0.3$.}
    \label{fig:QB-2}   
\end{figure}

The main characteristic feature of such collisions is the release of the $U(1)$ charge density. The charge, initially confined to the $Q$-ball, delocalizes in the form of Goldstone waves in the interior of the bubble, see Fig. \ref{fig:QB-1} and Fig. \ref{fig:QB-2}. As we previously observed, the bubble can be stabilized by Goldstone radiation or even grow significantly, see Fig. \ref{fig:QB-1}. Interestingly, the same happens in the so-called erasure processes, where a monopole or vortex collides with a wall-antiwall bubble \cite{Bachmaier:2023zmq, Dvali:2022rgx}. It was verified that the topological charge of a unit charge soliton dissolves inside the bubble. Both the topological and nontopological charges can be re-confined in a small spatial region if the bubble collapses. In higher dimensions, this happens in the case of a spherical bubble. Then, the localized solutions carrying a non-zero charge can reappear. So, a monopole/vortex \cite{Maxi-2}, as well as $Q$-ball/oscillon, with a nonzero value of the charge may be created.

However, there are two scenarios that have not yet been reported in the dynamics of topological solitons. First, in contrast to the unit charge monopole, the $Q$-ball may survive the collision with the bubble. Speaking precisely, {\it $Q$-balls can be emitted if the Goldstone wave hits the surface of the bubble}. This we clearly see in Fig. \ref{fig:QB-2} upper panels. This is possible because the $U(1)$ charge is not quantized and even its fraction can be confined inside a $Q$-ball emitted from the bubble surface. Nevertheless, this process may have a counterpart in the case of the monopole if the topological charge, and therefore the magnetic flux, is very large. Then, while a significant amount of the charge is released in the bulk of the wall-antiwall bubble, the rest could still escape in the form of a smaller charge monopole. This might reduce the efficiency of the erasure mechanism \cite{Dvali:1997sa}. 

Second, we observed a {\it transmutation process}, in which {\it the charge of the emitted $Q$ balls does not have the same sign as the absorbed $Q$ ball}, see Fig. \ref{fig:QB-2} upper panel. This is because the absorption of a $Q$-ball by the bubble excites the Goldstone waves with both negative and positive charge densities.

Third, it is also possible that not all the charge delocalizes. It happens when the $Q$-ball, after entering the bubble, transmutes into the oscillon in the false vacuum, see Fig. \ref{fig:QB-1} lower panels and Fig. \ref{fig:QB-2} upper panels. Although almost the whole charge is released in terms of the Goldstone waves, the false vacuum oscillon may still carry a small fraction of the charge density. Then, the oscillon propagates inside the bubble and can reach its boundary, where it finally disintegrates, Fig. \ref{fig:QB-2} upper panels. It is not excluded that a flux-carrying oscillon can be formed in the broken vacuum of a wall-antiwall bubble. Indeed, the abelian Higgs model is known to support an oscillon with a nontrivial flux density \cite{Gleiser:2007te}.

In some cases, especially in the thin-wall regime, the $Q$-ball feels a repulsive force from the bubble. Then, during the collision, only a part of the charge enters the interior of the bubble, see Fig. \ref{fig:QB-2} lower panels. The rest is backscattered in a charged oscillon as a remnant of the initial $Q$-ball.

%%%%%%%%%%%%%%%%%%%%%%%%%%%%%%
\section{Summary}
%%%%%%%%%%%%%%%%%%%%%%%%%%%%%%

In this paper, we systematically studied the ingredients that govern the dynamics of $Q$ balls. Following the kink and vortex dynamics, we identified two main factors: the vibrational modes and the ephemeral states, which are unstable, sometimes even short-lived states such as the polarized $Q$-balls (oscillons, charged oscillons, etc.), bubbles, and sphalerons. While (charged) oscillons  exist for any potential, the bubble and the sphaleron require an additional symmetry-breaking vacuum. 

Both factors have great impact on the $QQ^*$ collisions, leading to a huge complexity of the scattering scenarios. The vibrational modes trigger the resonant transfer of energy and lead to a chaotic, self-similar pattern in the final (and long-lived) state, which is qualitatively very similar to the kink-antikink case. This occurs in the thin-wall regime and is basically a potential independent feature. This should be contrasted with the thick-wall regime, where $Q$-balls do not host massive normal modes, and the collisions are much simpler. 

The complexity of the scenarios increases further if the potential possesses a false vacuum. Then, the typical intermediate state of the $QQ^*$ collision in the thin-wall regime is the formation of the false vacuum bubble. This makes collisions even more sensitive to the initial data. Importantly, the bubble is temporarily stabilized by the Goldstone waves. They show up due to the flat direction in the broken-symmetry vacuum and are formed from the release of the charge density originally localized on the $Q$-balls. The Goldstone waves exert pressure on the bubble boundaries and prevent it from collapse. Sometimes, a very long-lived state is formed with the Goldstone mode trapped in the bubble as a standing wave, i.e. a trapped normal mode. This gives a charge-swapping state of a new origin. In addition, the trapped Goldstone modes can trigger the spectral wall phenomenon, previously observed in domain wall and vortex dynamics. The false vacuum also supports a sphaleron (critical bubble), which can appear during the collision and, similar to kink scatterings \cite{Adam:2021fet}, modify the outcome. 

We note that the bubble is an example of a cavity-like bag solution. Interestingly, solutions of this type have recently been found in the context of oscillons in gapless models \cite{vanDissel:2025xqn}. The evolution of such oscillons resembles the evolution of the unperturbed bubble. Probably, the reported clean (dirty) regimes correspond to the bag with unexcited (excited) trapped modes. 

The dynamics of $Q$-balls can also be taken as a first step in understanding of dynamics of oscillons in multi-component models \cite{VanDissel:2020umg, Shafi:2024jig, Li:2025ioq}. This is because a $Q$-ball theory is equivalent to a model with two real scalars with the additional $U(1)$ target space symmetry. Such a symmetry is not present in a generic setup. Hence, generically, there are no stable time-dependent stationary solutions (which would correspond to $Q$-balls). Nevertheless, collisions of the oscillons would present a similar complexity with a similar importance of the modes and the ephemeral states. Of course, the dynamics of $Q$-balls in models with a more complicated target space \cite{Friedberg:1976me}, \cite{Alonso-Izquierdo:2023pti, Alonso-Izquierdo:2023xni, Alonso-Izquierdo:2024ohc} will be even more involved. 

 Of course, it would be very interesting to know how the interaction patterns are affected by quantum corrections, \cite{Dashen:1974ci, Goldstone:1974gf, Cahill:1976im, Faddeev:1977rm}. Here, the most promising approach seems to be the displacement operator framework, where the soliton is a squeezed coherent state \cite{Evslin:2019xte}. Indeed, several dynamical properties of quantum solitons have been computed, see e.g., \cite{Evslin:2022wyx, Evslin:2022fzf, Evslin:2023oub}. Importantly, it has been suggested that oscillons, which are the main actors in the $Q$-ball dynamics, may also be long-lived states at the quantum level \cite{Evslin:2023qbv}. 
 
 In summary, the $QQ^*$ collisions reveal a significantly more complicated structure than in the case of topological solitons in one and two spatial dimensions.

\section*{Acknowledgements}
D. C. M. and A. W. acknowledge support from the Spanish Ministerio de Ciencia e Innovacion (MCIN) with funding from the European Union NextGenerationEU (Grant No. PRTRC17.I1) and the Consejeria de Educacion from JCyL through the QCAYLE project, as well as the grant
PID2023-148409NB-I00 MTM.
K. S. acknowledges
financial support from the Polish National Science
Centre (Grant No. NCN 2021/43/D/ST2/01122). PMS and PED acknowledge partial support from  STFC Consolidated Grants ST/X000672/1 and ST/X000591/1 respectively.  A. W. acknowledges M. Bachmaier and S. Krusch for discussion and remarks.

\bibliographystyle{JHEP}
\bibliography{refs}

\end{document}